\documentclass[twocolumn,
aps,nofootinbib,showpacs,showkeys,preprint
tightenlines
] {revtex4}

\usepackage{epsf,epsfig,subfigure,graphicx,amsmath,amssymb}
\usepackage{color}
\newcommand{\dis}[1]{\begin{equation}\begin{split}#1\end{split}\end{equation}}

\newcommand{\GeV}{\,\mathrm{GeV}}

\newcommand{\pb}{\,\mathrm{pb}}
\newcommand{\invfb}{\,\mathrm{fb}^{-1}}

\def\etmiss{{E\hskip -0.2cm\slash \hskip 0.01cm }_T}
\def\bfetmiss{{\bf{E\hskip -0.2cm\slash \hskip 0.01cm }}_T}

\def\squark{\tilde{q}}
\def\gluino{\tilde{g}}
\def\chargino{\tilde{\chi}_1^{\pm}}
\def\2ndlnp{\tilde{\chi}_2^0}
\def\lnp{\tilde{\chi}_1^0}

\def\mmt{M_{T}}
\def\mt2{M_{T2}}
\def\trialmt2{M_{T2}(\chi)}

\def\mthetat2{M_{\theta T2}}
\def\trialmthetat2{M_{\theta T2}(\chi)}

\def\mc{M_C}
\def\mmct{M_{CT}}
\def\mct2{M_{CT2}}
\def\trialmct2{M_{CT2}(\chi)}

\def\mmctt{M_{CT,T}}
\def\mctt2{M_{CT2,T2}}

\def\mpit2{M_{\pi T2}}
\def\trialmpit2{M_{\pi T2}(\chi)}

\def\p0{{\bf p}_0}
\def\pzerot{{\bf p}_{0T}}

\def\etal{{\it et al.}}

\begin{document}


\title{\Large\bf Amplification of endpoint structure for new particle mass measurement at the LHC}

\author{Won Sang Cho\email{wscho@mulli.snu.ac.kr}}
\author{Jihn E. Kim\email{jekim@ctp.snu.ac.kr}}
\author{Ji-Hun Kim\email{jhkim@cpt.snu.ac.kr}} \affiliation{
Department of Physics and Astronomy and Center for Theoretical
Physics, Seoul National University, Seoul 151-747, Korea
 }
\begin{abstract}
We introduce a new collider variable, $\mct2$, named as
constransverse mass. It is a mixture of `stransverse mass($\mt2$)'
and `contransverse mass($\mmct$)' variables, where the usual
endpoint structure of $\mt2$ distribution can be amplified in the
$\mct2$ basis by large Jacobian factor which is controlled by trial
missing particle mass. Thus the $\mct2$ projection of events
increases our observability to measure several important endpoints
from new particle decays, which are usually expected to be buried by
irreducible backgrounds with various systematic uncertainties at the
LHC. In this paper we explain the phenomenology of endpoint
amplification in $\mct2$ projection, and describe how one may employ
this variable to measure several meaningful mass constraints of new
particles.
\end{abstract}


 \maketitle

\section{Introduction and motivation}\label{sec:Introduction}
The large hadron collider(LHC) will explore the TeV scale soon searching for new physics
beyond the Standard Model(SM) \cite{atlas,cms}, and various new
physics models are waiting to be tested.

A well-known expectation about the new physics phenomenology at the
LHC is that a $Z_2$ parity conservation could be a common feature in
some new physics models where the lightest new particle(LNP) are
very stable, providing weakly interacting massive particle(WIMP)
dark matter candidates. Supersymmetry(SUSY) with R-parity
\cite{susy}, Little Higgs models with T-parity \cite{littlehiggs},
or Universal Extra Dimension models with the
Kaluza-Klein parity \cite{extradim} are the examples of such models.
In those models, a pair of weakly interacting stable LNPs are
missing in the detector leaving rich missing transverse
energy($\etmiss$) signals. In general, the existence of multiple
missing particles makes an event
reconstruction very hard 
at the LHC where we will also suffer from the partonic center of
momentum(CM) frame unambiguity and complex new event topologies.
Under this circumstance, the mass measurement of new particles is
not an easy task at the LHC, and many important previous studies on
the subject might be useful
\cite{invmassedge,edgetail,massshell,mt,mttwo,mtmt2exp,mt2kink,
mt2kinkinclusive,
mt2kinkisr,mt2ext,mct,cusps,m2c,constmt2,mtgen,sqrtsmin,
momentaapprox,qcdisr,decompose}.

In particular, when a new event cannot be reconstructed, measuring
the endpoints of event projections onto various observables can be a
good way to obtain some mass constraints of the new particles
involved. It is because the endpoints correspond to the kinematic
boundaries of the allowed phase space of the event, which are mainly
described by the related particle masses. Some examples of such kind
of the methods include kinematic endpoint methods using invariant
masses of visible particles \cite{invmassedge}, and mass
measurements using the endpoints of various transverse mass
variables
\cite{mt,mttwo,mtmt2exp,mt2kink,mt2kinkinclusive,mt2kinkisr,mt2ext,mct}.
The existence of such endpoints or some cusp points in a
distribution can be re-analyzed by surveying singularity structures
of the allowed phase space of the events \cite{cusps}.

For these methods using endpoints, the precise measurement of the
endpoint must be the most important factor for it to be reliable.
However, identifying the correct endpoint is not an easy task in the
real situation since in general there exist complex systematic
uncertainties. Once we precisely understand the dynamics of the
signal/background processes and its detector responses, then the
mass parameters can be obtained via the least likelihood method
using templates as in Ref. \cite{mtmt2exp}. This
method is reliable because the Standard Model(SM) has been
well-understood, explaining most of the phenomena in the high energy
experiments up to now. However, when it comes to the LHC, things
will be changed. If there exist new physics beyond the SM at the TeV
scale, it is quite obvious that we will suffer from huge systematic
uncertainties in measuring the masses or new model parameters. For
example, about the signal with jets, heavy jet combinatoric
backgrounds should be understood in the situation where the new
physics events can have very complex event topologies
with hard QCD effects on the beyond the SM(BSM) signals.

In this regard, extracting the meaningful endpoints via some
simplified functional fitting have been performed near a proper and
plausible endpoint region of the distribution
\cite{invmassedge,mt,mttwo,mtmt2exp,mt2kink,mt2ext,mct,
mt2kinkinclusive,mt2kinkisr}.
In that region, a Gaussian smeared segmented linear functional
fitting for signal and backgrounds is likely to give a sufficiently
good description in many cases. However, it is true that one would
still suffer from large systematic uncertainties in these segmented
straight-line fits to extract the endpoint position. In particular,
it is more severe when the signal endpoint structure is faint with
feet or tails (ignoring any smearing effects from finite total decay
width or detector resolutions), leaving small number of events near
the endpoints, not constructing a sharp drop \cite{edgetail}. In
many cases, the endpoint fitting has quite a large uncertainty with
irreducible backgrounds.

In this paper, we assume a situation where one suffers from heavy
systematic uncertainties in identifying the endpoints, and define
the meaningful signal endpoint as a {\it breakpoint(BP) in the
distribution up to smearing effects}. We describe a way to obtain
some mass constraints more precisely, which have been obtained from
the endpoints of $\mt2$ distributions. To do so, we introduce a new
variable, $\mct2$, named as `constransverse mass'. The $\mct2$
projection of the events is found to have interesting Jacobian
factor with respect to $\mt2$ distribution, which amplifies the
endpoint structures of $\mt2$.

In Sec. II, we start with the definition of $\mmct$ which is the
basic ingredient for the $\mct2$ study, and describe the
phenomenology of the endpoint structure amplification, comparing it
with the $\mmt$ distribution. In Sec. \ref{sec:mct2}, the $\mct2$
variable is introduced. In Sec.\ref{sec:mct2experiment} it is
described how one may employ the $\mct2$ projection for a precise
measurement of the signal endpoint, usually buried in the
irreducible jet combinatoric backgrounds. Sec.\ref{sec:Conclusion}
is a conclusion of our study.

\section{Properties of $\mmct$}\label{sec:mct}
Let us consider a mother particle $Y$ decaying to the visible and
missing particles ($Y \rightarrow v(p) + X(k) $), where $p$ and $k$
are their corresponding 4 momenta. Then, one can reconstruct
transverse mass($M_T$) \cite{mt} which is bounded from above by the
invariant mass($M$) of $Y$, \dis{ M^2&\equiv
m_v^2 + m_{X}^2 + 2 (E_v E_{X} - \vec{p}\cdot \vec{k})  \\
&=m_v^2 + m_{X}^2 + 2 (e_v e_{X} \cosh(\Delta \eta) - \vec{p}_T\cdot \vec{k}_T) \\
&\geq M_T^2 \label{eqinvm} } where  $e_v=\sqrt{m_v^2+|\vec{p}_T|}$
and $e_{X}=\sqrt{m_{X}^2+|\vec{k}_T|^2}$ are transverse energies of
the visible and missing particles with their transverse momenta,
$\vec{p}_T$ and $\vec{k}_T$, $\Delta \eta$ is the rapidity
difference between $v$ and $X$, and $M_T^2$ is defined by
\begin{eqnarray}
M_T^2&\equiv& m_v^2 + m_{X}^2 + 2 (e_v e_{X} - \vec{p}_T\cdot
\vec{k}_T). \label{eqMTsquare}
\end{eqnarray}
Furthermore, we can also define $\mc$ and $\mmct$ in terms of
$v$ and $X$ \cite{mct},
\dis{
\mc^2&\equiv m_v^2 + m_{X}^2 + 2 A, \\
\mmct^2 &\equiv m_v^2 + m_{X}^2 +2 A_T,\\
A&\equiv E_v E_{X} + \vec{p}\cdot \vec{k},\\
A_T&\equiv e_v e_{X} + \vec{p}_T\cdot \vec{k}_T,\label{eqAterms} }
satisfying a similar inequality,
\begin{eqnarray}
\mc^2\geq \mmct^2~.
\end{eqnarray}
An interesting property of $\mc$
variable is that it is invariant under the back-to-back boost of the
two particles,
\begin{eqnarray}
&&p^\mu\rightarrow \Lambda^\mu_\nu(\vec{\beta})p^\nu,\quad
k^\mu\rightarrow \Lambda^\mu_\nu(-\vec{\beta})k^\nu,
\label{b2bboost}
\end{eqnarray}
where $\Lambda^\mu_\nu(\vec{\beta})$ denotes the Lorentz
transformation matrix for the boost parameter $\vec{\beta}$. The
back-to-back boost invariance(BBI) results from the BBI of
``A-term", of (\ref{eqAterms}), which is the Euclidean dot products
of $p$ and $k$. The BBI of the Euclidean momenta product has been
noticed in \cite{mt2kink,mtgen} with the form of $A_T$ in $\mt2$
solutions, and utilized in $\mmct$ \cite{mct} using the visible
transverse momenta.

In the rest frame of $Y$, $\mc$ and $\mmct$ read
\dis{
&\mc^2= m_v^2 + m_{X}^2 + 2 (e_v^0 e_{X}^0\cosh(\Sigma\eta^0)
-|\pzerot|^2) \\
&\mmct^2= m_v^2 + m_{X}^2 + 2 (e_v^0 e_{X}^0 - |\pzerot|^2)
\label{eqmc2} } where $e_{v,X}^0$ are the transverse energies of the
daughter particles, $\Sigma \eta^0$ is the rapidity sum of $v$ and
$X$, and
$|\p0|=\lambda^{1/2}(m_Y^2,m_X^2,m_v^2)/2m_Y$,\footnote{$\lambda(x,y,z)\equiv
x^2 + y^2 +z^2 - 2 (xy + yz + xz)$} which is the absolute momentum
of daughter particles \textit{in the $Y$ rest frame}. The reason
that we specify $\mc$ and $\mmct$ in the rest frame of $Y$ as in
(\ref{eqmc2}), is because $\mc$ is not a frame independent quantity.
\textit{From now on, we will use the value of
$\mc$ defined only in the rest frame of mother particle as in
(\ref{eqmc2}), however we will not require such a restriction on
$\mmct$.}

The distribution of $\mmct$ is interesting. In particular, the range
of the distribution can be drastically changed with respect to the
input trial missing particle mass $\chi$. Let us assume that the
true value of missing particle mass is not known. We can also
construct $\mmt(\chi)$ and $\mmct(\chi)$ using a trial invisible
mass, $\chi$, by replacing $m_X$ to $\chi$ in
(\ref{eqinvm},\ref{eqMTsquare},\ref{eqmc2}).
Then, as long as the $Y$ is transversely at rest in the
lab. frame, $\mmct(\chi)$ is bounded from above by $\mc(\chi)$ as
follows \dis{
\mc^2&= \chi^2 + 2 (|\p0|E_{X}^0 -|\p0|^2), \\
 \mmct^2&= \chi^2 + 2 (|\pzerot| e_{X}^0 - |\pzerot|^2),\\
&\quad\mc^2\geq  \mmct^2 \label{eqmc3}
}
in which $M_T(\chi)$ also satisfies a similar relation,
\dis{
M^2&= \chi^2 + 2 (|\p0|E_{X}^0 +|\p0|^2)  \\
 M_{T}^2&= \chi^2 + 2 (|\pzerot| e_{X}^0 +|\pzerot|^2),\\
&\quad M^2\geq M_T^2\label{eqmc4} }
 where $E_X(e_X)=\sqrt{\chi^2 +
|\p0\,(\pzerot)|^2},$ and the visible masses are ignored in both
cases. It is very well known that $\mmt(\chi=m_X)$ has a
nice endpoint, $m_Y$, which is invariant under the transverse boost
of $Y$, even though the bulk distribution is shifted by $P_T$ of
$Y$. However, when $\chi$ is not the true value (the maximum of
$\mmt(\chi)$), $M(\chi)$ is not invariant any more under the
transverse motion of the $Y$.\footnote{The features of
transversely boosted $\mmt(\chi)$ and $\mt2(\chi)$ endpoints were
surveyed by A. Barr et. al. in \cite{mt2kink} and M. Burns et. al.
in \cite{mt2ext}} It is also true for the maximum of $\mmct(\chi\neq
m_{X})$. In the following section, we will comment also on the feature of the shift of $\mmct(\chi)$ endpoint.

Then, let us compare the distributions of $\mmct(\chi)$
and $\mmt(\chi)$ when the mother particle is transversely at rest in
the lab. frame. From Eqs. (\ref{eqmc3},\ref{eqmc4}), we note that $\mmct(\chi)$ and $M_T(\chi)$ share the same minimum point at $\chi$ but get
different maxima. As $\chi$ tends to zero, $\mc(\chi)$ tends zero,
and $\mmct(\chi)$ distribution collapses almost to a zero point,
while $M_T$ ranges from zero to $2|\p0|$. On the other hand, when
$\chi$ is very large compared to $|\p0|$, their ranges and shapes
become very similar because the effect of the sign flip in the
squared momentum is negligible. The most impressive property of the
$\mmct(\chi)$ projection is that if we use a proper value of
$\hat{\chi}(\equiv \chi/|\p0|)$, which is not quite large, then the
endpoint structure of $\mmct(\chi)$ distribution can be enhanced by
some large Jacobian factor $J_{max}(\hat{\chi})$ compared to the
$M_T(\chi)$ endpoint distribution. The Jacobian factor is given by
\begin{displaymath}
\sigma^{-1}\frac{d\sigma}{d\mmct(\chi)}\sim
J\sigma^{-1}\frac{d\sigma}{d\mmt(\chi)},
\end{displaymath}
or
\begin{eqnarray}
 J =\frac{\mmct(\chi)}{M_T(\chi)}\frac{(e_{X}+|\pzerot|)^2
 }{(e_{X}-|\pzerot|)^2}\label{eqjacobian}
\end{eqnarray}
whose behavior is
\begin{displaymath}
J\rightarrow \left\{
\begin{array}{ll}
\frac{\mc(\chi)}{M(\chi)}\frac{(E_{X}+|\p0|)^2}{(E_{X}-|\p0|)^2}=J_{max}, & \textrm{ the endpoint region, }\\
\\
\quad\quad 1, & \textrm{ the minimum region.}
\end{array} \right.\
\end{displaymath}
Thus, the endpoint enhancement factor $J_{max}$ approaches $\infty$
when $\hat{\chi} (\equiv \chi/|\p0|)$ goes to zero, or becomes one
if $\hat{\chi}$ is very large. As a result of the very different
compression rate between the endpoint and minimum region of $\mmt$,
most of the large $\mmt$ events are accumulated in the narrow
maximal region of $\mmct$ distribution if $\hat{\chi}$ is not so
large.
\begin{figure}[ht]
\begin{center}
\epsfig{figure=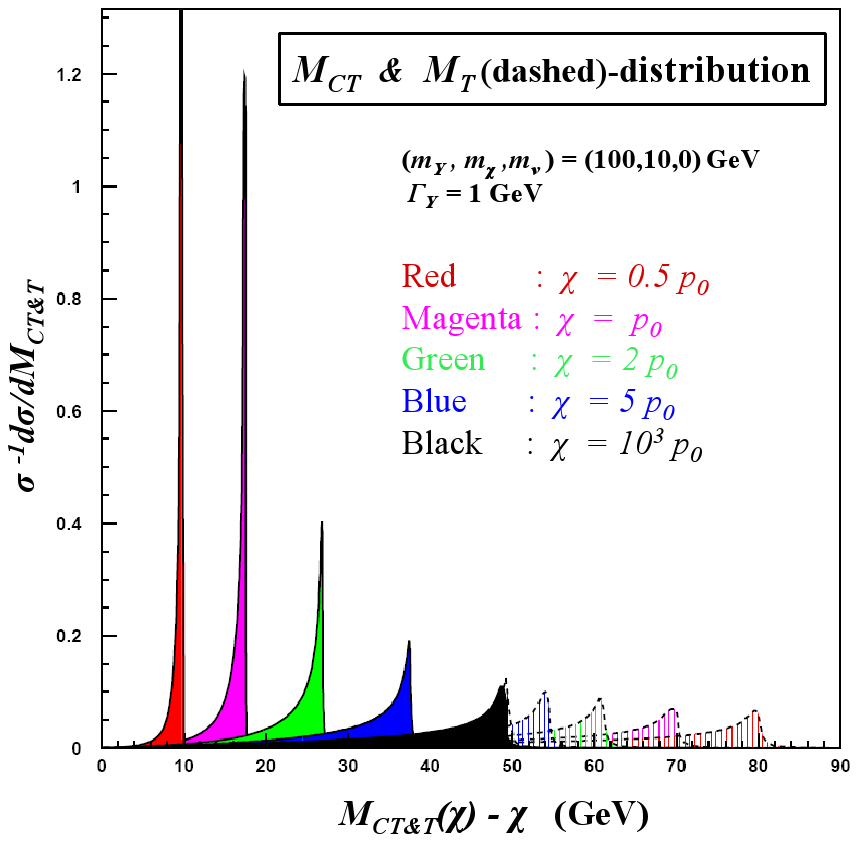,width=8cm,height=8cm,angle=0}
\epsfig{figure=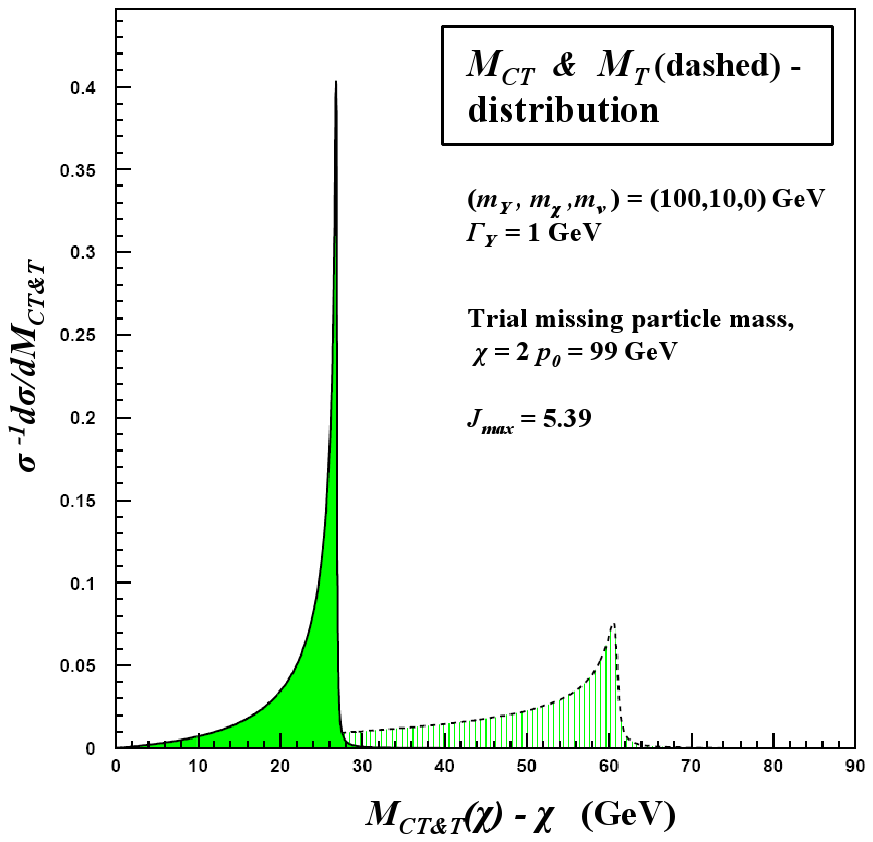,width=8cm,height=8cm,angle=0}
\caption{Enhancement at the  endpoint.  } \label{fig:mtmct}
\end{center}
\end{figure}
Fig. \ref{fig:mtmct} shows the distribution of $\mmct(\chi)$ and
$\mmt(\chi)$ when $Y$ is transversely at rest in the
lab. frame with $m_Y = 100 \,\GeV$, $m_\chi = 10 \,\GeV$, and $m_v
=0 \,\GeV$. The total decay width of $Y$, $\Gamma_Y$, was assumed to
be 1 $\GeV$ and spin angular correlation effect was ignored. In Fig.
\ref{fig:mtmct}(a), we used a trial missing particle mass $\chi$
corresponding to $2 |\p0|(=99 \GeV)$ where $J_{max}$ is $5.39$. The
black dashed line filled with green hatch corresponds to
$\mmt(\chi)-\chi$ distribution and the black line is
$\mmct(\chi)-\chi$ distribution. As explained, $\mmct$ and $\mmt$
share the same minimum at $\chi$, but the endpoint of $\mmct$
shrinks into a lower region so that the expected endpoint is
$\mmct^{max}(99)-99 = 26.9 \GeV$ while $\mmt^{max}(99)-99 = 61.2
\GeV$, ignoring the width effect. The height of the peak in $\mmct$
distribution is amplified by the $J_{max}$ factor compared to the
peak of $\mmt$. Fig. \ref{fig:mtmct}(b) shows how $\mmct$ and $\mmt$
are varied with respect to the trial missing particle mass $\chi$.
As shown in the figure, when the $\chi$ is small, $J_{max}$
diverges(red: $J_{max}=105$), and $\mmct$ distribution collapses
into a narrow peak region. On the other hand, if $\chi$ is large,
then $\mmct$ and $\mmt$ become the same (black: $J\sim1$).

Now we have a collider variable $\mmct(\chi)$ which can drastically
change and magnify its endpoint structure compared to that of
$\mmt(\chi)$. Basically, they provide the same physical constraint.
Nevertheless, there is a big difference on the systematic uncertainties at these two end point regions.
 So, we move on to discuss how the amplification of the endpoint
 structure accompanies the propagation of statistical and systematic uncertainties.

In principle, one can try to use the sharp amplified $\mmct(\chi)$
endpoint instead of the $\mmt(\chi)$ endpoint in order to measure
the mass constraint $|\p0|$ more precisely. However, one must be
sure about that a naive statistical uncertainty cannot be reduced
since as $\delta \mmct^{max}(\chi)$ decreases to $\frac{1}{J_{max}}
\delta \mmt^{max}(\chi)$, the error propagation factor
for $|\p0|$ increase by $J_{max}$. On the other
hands, things change when it comes to the systematic uncertainties.
Actually, the endpoint structures we want to resolve are not that
simple. If we attempt an accurate mass(endpoint) measurement with
exact distribution shapes or templates, one have to understand the
important dynamics of the new signal and backgrounds(surely also
with the good understanding of the detector  responses), which must
include large systematic uncertainties.
In this regard, model independent mass(endpoint) measurement will be
important.

An effective way to measure the endpoint is simply observing and
pinpointing the breakpoint(BP) with simplified local
fit function up to some smearing width in the distribution. In this
respect, it is important to know the effective range
near the BP for the simplified local fit functions to
be reliable, because we suffer from a large uncertainty in the bulk
distribution. However, it is not apparent when the signal endpoint
structure is dim and faint with feet or long tail above some
irreducible backgrounds so that the BP is quite ambiguous with a
small slope discontinuity. [We will assume that the background
distribution is not singular at the BP.] When we attempt to identify
a BP as the position of a meaningful endpoint, a slope discontinuity
between the lower and the upper region of the BP is crucial factor
for the BP measurement with less errors. The propagated uncertainty
of the BP can be given as follows:
\begin{eqnarray}
\delta_{BP}^2 \sim \frac{\sigma^2}{\Delta a^2}, \label{eqbperror}
\end{eqnarray}
where $\sigma$ is an error which can be caused by various
sources, and $\Delta a$ is the slope difference between the two
regions segmented by the BP. $\sigma$ may come from event
statistics or some systematic uncertainties. The point we note here
is that the larger the slope difference, the clearer the endpoint
structure, and as a result it enables us to elaborate the fit function
and to choose more effective range of the fit function. In this regard, the
elaboration of the fit scheme will reduce $\sigma^{sys}$, and as a result
$\delta_{BP}^{sys}$ will be reduced.

For the same reason an ambiguous faint endpoint in the $\mmt$
distribution can be measured more precisely in the $\mmct$ projection, and as a result the systematic error in obtaining the mass constraint
$|\p0|$, $\delta_{\p0}^{sys}$, can be reduced by $O(1/J_{max})$. The specific estimation of error suppression factor is based
on the following two facts,
\begin{itemize}
\item[(1)] The $\mmct(\chi)$ projection amplifies the slope
difference $\Delta a$ near the $\mmt(\chi)$ endpoint by $J_{max}^2$,
\begin{eqnarray}
\Delta a' \rightarrow J_{max}^2(\chi) \Delta a
\end{eqnarray}
\item[(2)] By the enhancement of the slope discontinuity,
$\sigma^{sys}$ in the $\mmct$ projection can be less than, or at
least comparable to the $\mmt$ case with refined fit scheme near the
enhanced endpoint
\begin{eqnarray}
\sigma'^{sys} \sim \sigma^{sys}~ .
\end{eqnarray}
\end{itemize}
Then $\delta^{sys}_{\mmct^{max}} \rightarrow
\frac{1}{J_{max}^2}\delta_{\mmt^{max}}^{sys}$, and if we take into
account the error propagation factor $J_{max}$ for $|\p0|$, we get
$\delta_{\p0}^{sys} \rightarrow
\frac{1}{J_{max}}\delta_{\p0}^{sys}$.

Up to now we described that $\mmct(\chi)$ is well-defined variable
using $\vec{p}_T$ and trial missing particle mass $\chi$ in the rest
frame of $Y$. The large amplification factor $J_{max}$ can be
obtained simply by changing $\chi$, and there is no additional
parameter needed to accentuate the dim BP of the
$\mmt(\chi)$ distribution. One might have a good chance to reduce
some systematic uncertainty in extracting the mass constraints with
the $\mmct$ projection.

\section{constransverse mass, $\mct2$}\label{sec:mct2}
In this section we introduce constransverse
mass($\mct2$) for new physics events with two missing LNPs,
which inherits the properties of $\mmct$. Some differences of
using $\mmct$ and $\mct2$ will be discussed also.

When can we observe such an $\mmct$ endpoint in a real experiment?
One thing we should remind is that the maximum of
$\mmct(\chi)$ and $\mmt(\chi)$, $\mmct^{max}(\chi)$ and
$\mmt^{max}(\chi)$, are not invariant under the mother particle's
transverse motion when $\chi$ is not true value. One more
shortcoming for $\mmct^{max}(\chi)$ is that it is not also
invariant under the transverse boost of the mother particle, even
when  $\chi$ is true value.
The reason for this is simply that $\mc$ in (\ref{eqAterms}) is
not an invariant quantity under the boost of mother particle, while
 $M(m_{\chi})$ (the endpoint value of $\mmt(m_{\chi})$) is an
invariant one, as $\mmt^{max}$ has been utilized to measure the
W-boson mass provided with already known neutrino mass
\cite{mt,mtmt2exp}. However if we assume that we do not know the mass
of the missing LNP, $m_{\chi}$, while we just want to measure some
mass constraints related to the decay process, then
$\mmct^{max}(\chi)$ has no disadvantages compared to
$\mmt^{max}(\chi)$. As we will see in the examples of the following section,
it can be much better to use $\mmct^{max}(\chi)$ or
$\mct2^{max}(\chi)$ for this purpose. Then, it is just required for
us to choose our decay system of interest being at rest in the
transverse direction.

When the mother particle is boosted in the transverse direction  $x$
with momentum, $\delta_T$, the shift of the $\mmct(m_{\chi})$
endpoint is described by
\begin{displaymath}
\frac{\Delta\mmct^{max}(m_{\chi})}{\mc(m_{\chi})} \sim
f(\hat{m_{\chi}})\,\alpha \cos\phi, \,\,\,\textrm{$(\alpha \ll 1)$}
\end{displaymath}
where $f(\hat{m}_{\chi})=\frac{4(\hat{E}_{X}
-1)}{\hat{m}_{\chi}^2+2(\hat{E}_{X}-1)}$,
$(\hat{m}_{\chi},\hat{E}_{X})=(\frac{m_{\chi}}{|\p0|},\frac{E_{X}}{|\p0|})$,
$\alpha = \delta_T/m_{Y}$, and $\phi$ is the azimuthal angle of
visible particle in the mother particle's rest frame.
The shift is of order $\alpha$ in general, and
one should take care of the use of the $\mmct(\chi)$ endpoint for
the mother particle with sizable transverse momentum in the lab.
frame. $\delta_{T}^{max}$ cut must be accompanied with a proper
event selection. Surely, the $\delta_{T}$ suppression cut might
reduce the statistics of the signal event candidates. However, the
cut can also play a role for purifying the signatures from
backgrounds in many cases. We will describe such an effect of the
cut with a new physics example, using $\mct2$ endpoint in the next
section.

Then, let us consider the production of a pair of identical new physics
particles at the LHC, in which each of two mother particles
$Y_i$ decays to visible $V_i$ and invisible particles $X_i$, ($p + p
\rightarrow G(-\delta_T) + Y_{i}( \rightarrow v_{i}(p_{i}) +
X_{i}(k_{i})), i=1,2 $). Here $G$ with its transverse
momentum($-\delta_T$) denotes the other particles not from the $Y_i$
decays, so that it provides the transverse momenta of $Y_1 + Y_2$
system as $\delta_T$. In this case, we find that it is also possible
to observe the same $\mmct(\chi)$-like endpoint by the construction
of \textit{constransverse mass}, $\mct2(\chi)$ by
\begin{eqnarray}
\mct2(\chi) &=&\min_{\sum{{\bf k}_{iT}}=\bfetmiss}
[\max{\{\mmct^{(1)},\mmct^{(2)}\}}]\label{eqmct2}\\
\mmct^{(i=1,2)}(\chi)^{2} &=& \chi^2 + m_{v_i}^2 + 2 (e_{v_i}
e_{\chi_{i}} + \vec{{\bf p}}_{iT} \cdot \vec{{\bf k}}_{iT}),
\nonumber
\end{eqnarray}
where $\chi$ is a trial missing particle mass, and
$\bfetmiss=-\sum{{\bf p}_{iT}} + \delta_T$. Here all the particle
momenta are defined in the lab. frame. $\mmct^{(i)}(\chi)$ is the
contransverse mass of each mother particle system as defined in
(\ref{eqAterms}) using a trial transverse momenta ${\bf k}_{iT}$
which satisfies the missing transverse momentum condition. This is
the contra version of Cambridge stransverse mass ($\mt2$) variable
\cite{mttwo}. Please note that $\mmct^{(i)}$ in (\ref{eqmct2}) is
defined \textit{for the mother particle pair where each
mother particle can have sizable transverse momentum in the lab.
frame}, and it can make $\mmct^{(i)max}(\chi)$ shifted as explained
previously. However, an interesting point is that if $\delta_T = 0$,
$Y_1+Y_2$ system is transversely at rest, and the
$\mct2^{max}(\chi)$ also has well-defined endpoint as $\mc(\chi)$ in
(\ref{eqmc3}). This is analogous to the case that $\mmct^{max}(\chi)$ becomes $\mc(\chi)$ when the mother particle $Y$ has no transverse
momentum. To be analytically exact for that result, let us see the
solution of $\mct2(\chi)$. If $\delta_T = 0$, the analytic
$\mct2(\chi)$ solution is given as follows,
\begin{eqnarray}
\mct2^2(\chi)&=& \chi^2 - A_{T} + \sqrt{A_T^2+ 2 A_T
\chi^2},\label{eqmct2sol}\\ A_{T} &=& |{\bf p}_{1T}||{\bf
p}_{2T}|+{\bf p}_{1T}\cdot{\bf p}_{2T}, \nonumber
\end{eqnarray}
ignoring the mass of the visible particles.\footnote{In the rest of
this paper, we concentrate on the events with one massless visible
particle in each decay chain.}
The solution can be
obtained just by using the visible momenta with the sign flip in solving
the balanced equation as in Refs. \cite{mt2kink,mtgen}. From
(\ref{eqmct2sol}), one can see that
$\mct2^{max}(\chi)=\mmct^{max}(\chi)$ as the natural extension of
$\mmct$,
 \dis{
\mct2^{max}(\chi) &= \mc(\chi) \\
&= \chi^2 + 2 (|\p0|E_{X}^0 -|\p0|^2),\label{eqmct2max} }
in the limit of $\delta_T = 0$. This can be easily proved from Eq.
(\ref{eqmct2sol}), assuming the situation where the two mother
particles are at rest in the lab. frame and all the relic particles
are in the transverse plane. Such a pair of rest mother particles
can be guaranteed by the back-to-back boost invariance of
$\mct2(\chi)$ solution (\ref{eqmct2sol}). It is interesting to note
that $\mct2^{max}(\chi)$ is described by the $\p0$ value in
general, in spite of the nonzero transverse momentum for each of the mother particles. This is the same phenomenon that  $\mt2$
inherits most of the important properties of $\mmt$, so that
$\mt2^{max}(\chi)=\mmt^{max}(\chi)$ also. Physically Eq.
(\ref{eqmct2max}) can be satisfied by the events with kinematically
identical decay chains, each of which has the maximal $\mmt$ or $\mmct$
configuration. So in this case, $\mt2$ (or $\mct2$) can be
effectively the same as the single $\mmt$(or $\mmct$). This is also true
for the event contributing to the minimum.

The other important inheritance of $\mct2(\chi)$ from $\mmct(\chi)$
is that the Jacobian factor in (\ref{eqjacobian}) should be valid
also between the $\mct2(\chi)$ and $\mt2(\chi)$, at least for the
maximum and minimum regions of both variables as explained. In the
maximum region, the $\mct2(\chi)$ and $\mt2(\chi)$ solutions are
perfectly described by $\mc(\chi)$ in (\ref{eqmct2max}) and
$M(\chi)$ in (\ref{eqmc4}), respectively, while in the minimum
region, both $\mct2(\chi)$ and $\mt2(\chi)$ become the same value,
$\chi$ giving the Jacobian factor of $1$.

\begin{figure}[t]
\begin{center}%
\epsfig{figure=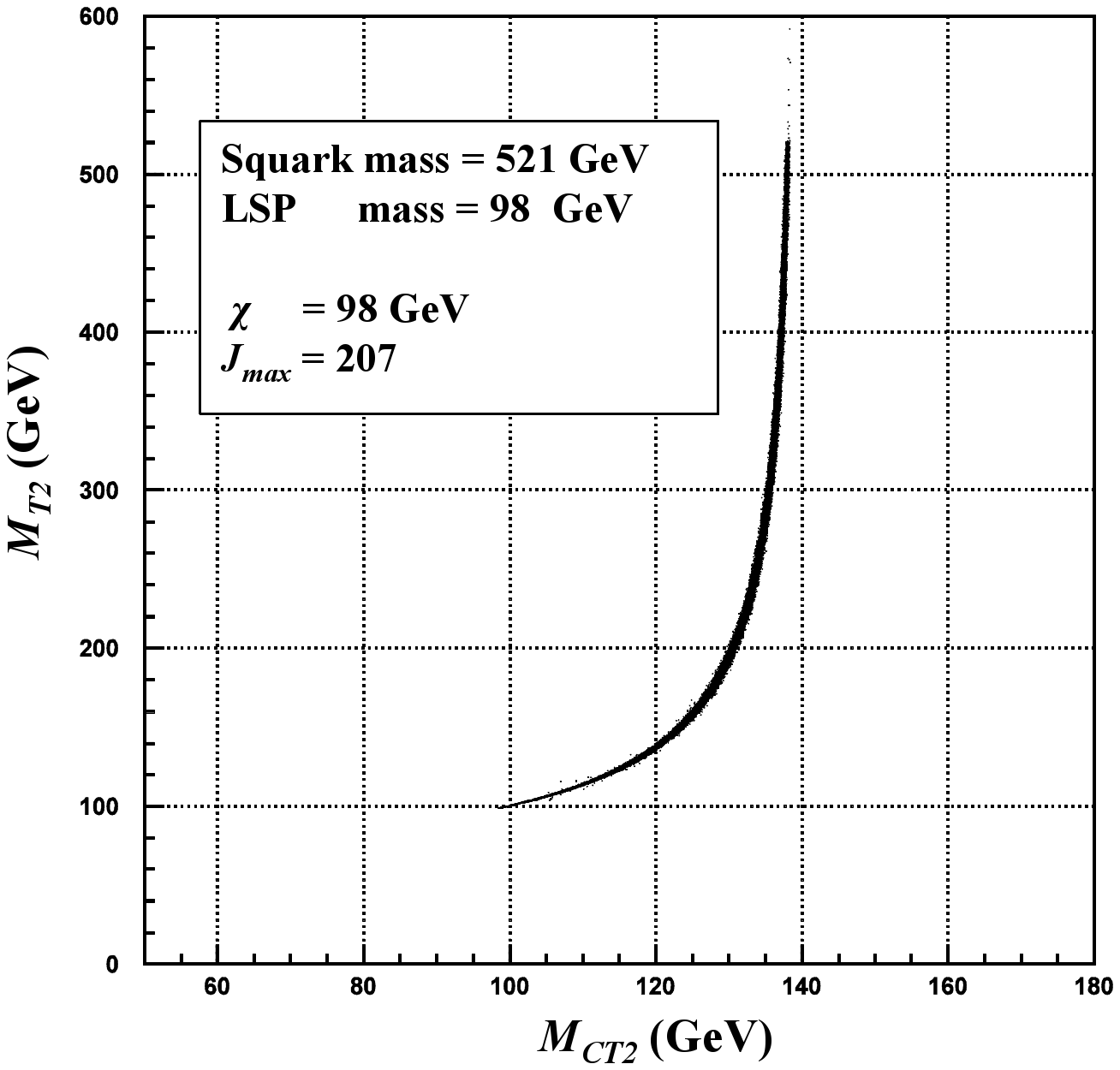,width=8.5cm,height=8cm,angle=0}
\epsfig{figure=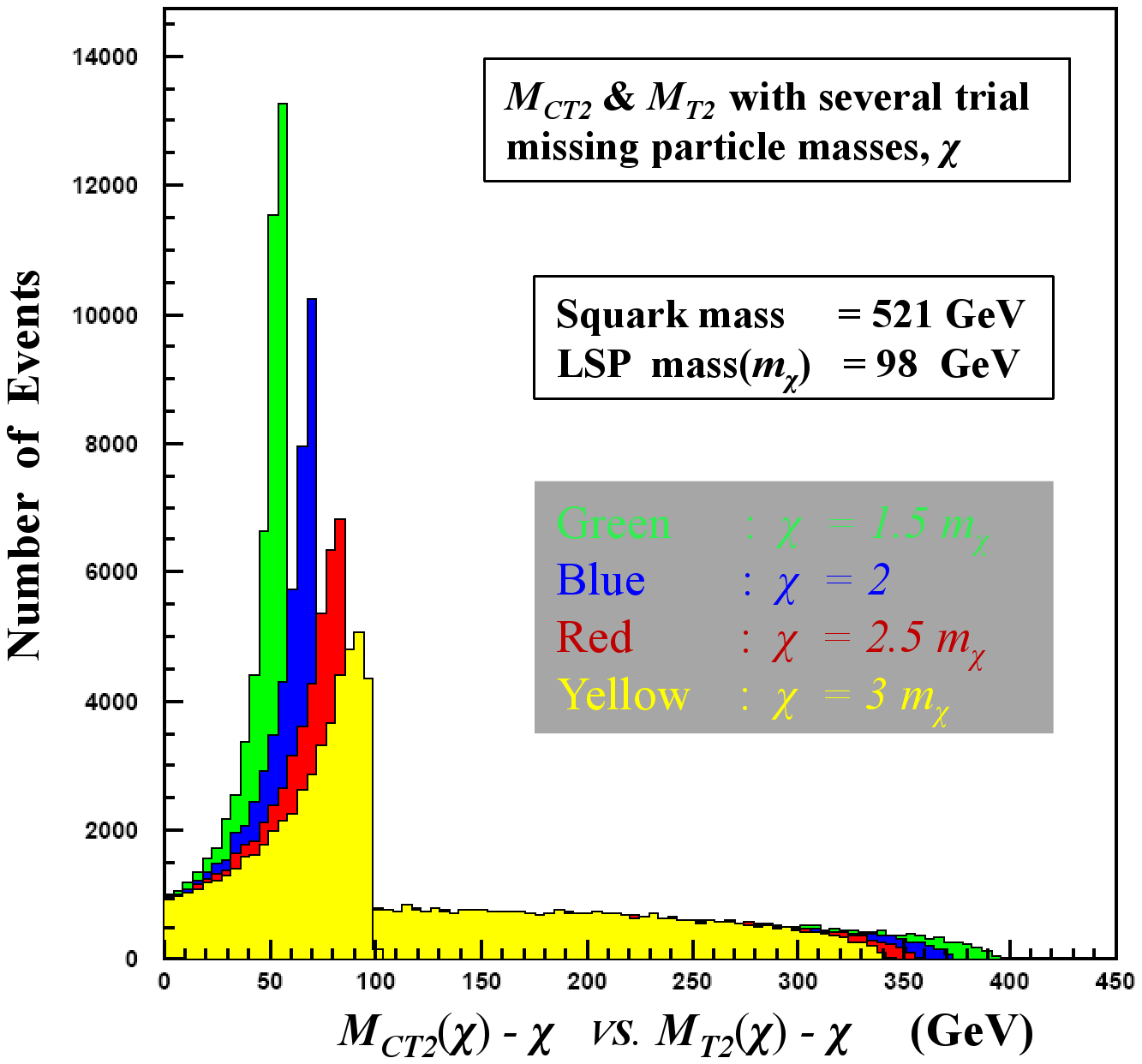,width=8.5cm,height=8cm,angle=0}
\caption{(a) $\mt2$ vs. $\mct2$ of SUSY(sps1a)
$\tilde{q}_R(\rightarrow q + \tilde{\chi}_{1}^{0})$ pair production
with a trial WIMP mass of 100 GeV, and (b) $\mct2(\chi)-\chi$
(uprisen) and $\mt2(\chi)-\chi$ distributions for $\chi =
1.5m_{\chi}$[green], $2m_{\chi}$[blue], $2.5m_{\chi}$[red], and
$3m_{\chi}$[yellow]. }\label{fig:mt2mct2}
\end{center}
\end{figure}

Fig. \ref{fig:mt2mct2} shows how the previous analysis of $\mmt$ and
$\mmct$ is extended to the $\mt2$ and $\mct2$ analysis for the SUSY(sps1a) event
when a pair of right handed squarks($m_{\tilde{q}_R}=521\GeV$) are
produced, and each decays to a quark and an
LSP($m_{\chi}=98\GeV$),($pp\rightarrow\tilde{q}_R\tilde{q}_R
\rightarrow 2\times(q+\tilde{\chi}_1^0)$). It is a parton level
simulation with no initial state radiation(ISR) effect so that $\delta_T \sim 0$.

Fig. \ref{fig:mt2mct2}(a) is a scatter plot of the events in the
$\mt2$-$\mct2$ plane for $\chi=m_{\chi}$. As expected from the
previous $\mmct$ and $\mmt$ analysis, most of the large $\mt2$
events are projected into a narrow endpoint region of the $\mct2$
variable, reflecting the large $J_{max}$ factor for the endpoint
region of $\mct2$ distribution. Fig. \ref{fig:mt2mct2}(b) is a
comparison plot between the  $\mt2(\chi)$ and $\mct2(\chi)$
distributions for $\chi = 1.5m_{\chi}$[green], $2m_{\chi}$[blue],
$2.5m_{\chi}$[red], and $3m_{\chi}$[yellow]. The uprisen
distributions are $\mct2(\chi)-\chi$ and the laid distributions are
$\mt2(\chi)-\chi$. For $\chi = (1.5\sim 3) m_{\chi}$, the
endpoint enhancement factors, $J_{max}(\chi)$ of (\ref{eqjacobian}),
are estimated to be $69$, $33$, $20$, and $13$. As in Fig.
\ref{fig:mtmct}, we can also observe a similar behavior such that if
$\chi$ is small then the $\mct2(\chi)$ distribution collapses into a
region near zero with a diverging $J_{max}$, while it approaches to
$\mt2(\chi)$ when $\chi$ has a large value.

\begin{figure}[ht!]
 \begin{center}
 \epsfig{figure=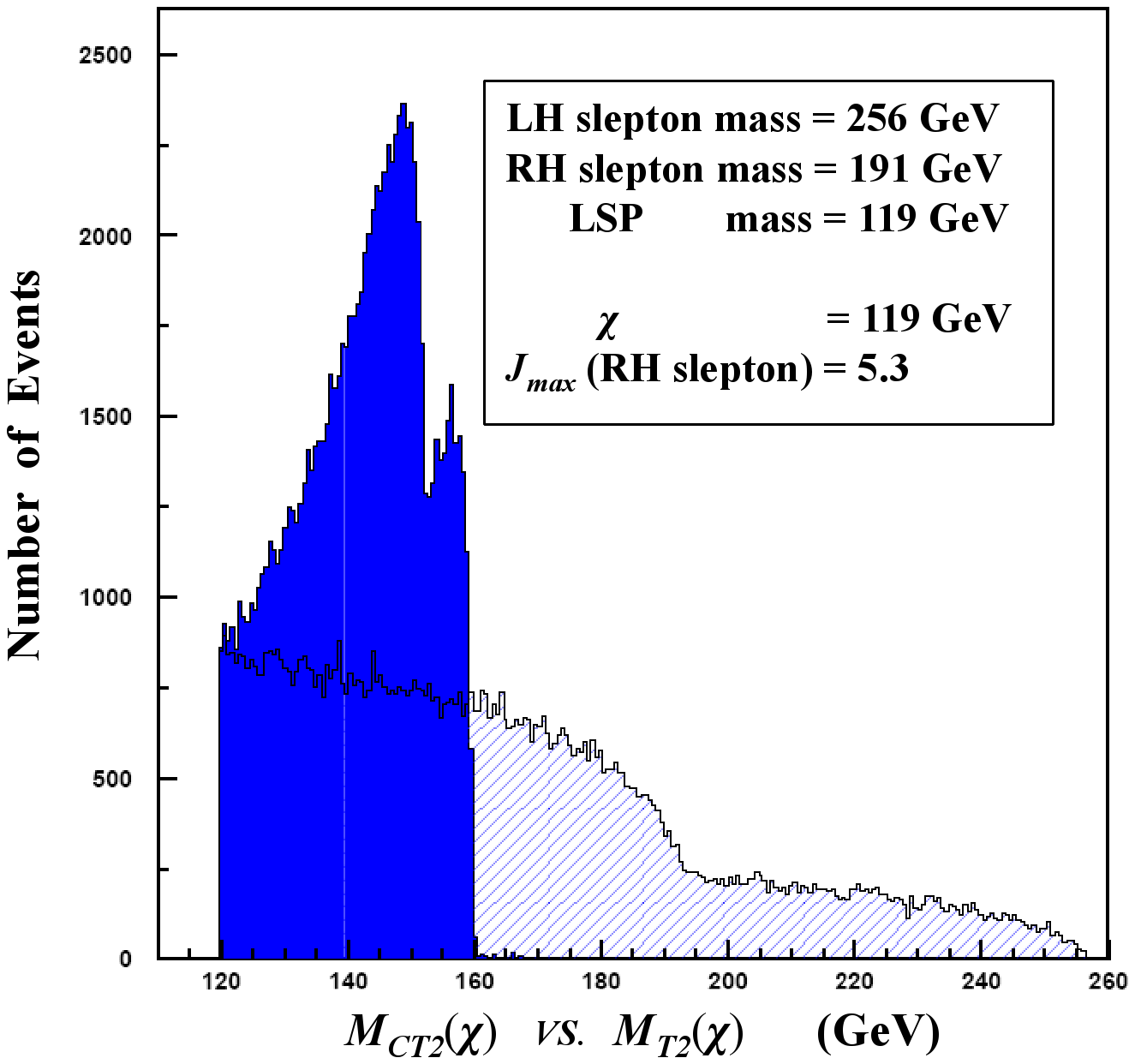,width=8.5cm,height=8cm,angle=0}
\caption{ The amplified endpoint structure of the RH and LH sleptons
in $\mct2$(uprisen) distribution and usual $\mt2(\chi)$
distribution(laid)} \label{fig:mt2mct2_uprising}
\end{center}
\end{figure}

As a result of a large $J_{max}$ factor, a small slope difference in
the $\mt2(\chi)$ distribution can be amplified also in the
$\mct2(\chi)$ projection by a factor of $J_{max}^2$, naturally
transformed into a remarkably enhanced or uprisen endpoint structure
of the mother particles. Therefore, as discussed in the previous
section, the $\mct2(\chi)$ projection also can help us to extract a
meaningful endpoint defined as a BP in the $\mt2(\chi)$
distribution. Even though there exist heavy systematic uncertainties
in the signal and backgrounds, the dim BPs will be amplified and it
can significantly reduce the systematic uncertainties we suffered
from in fitting with more elaborated fitting schemes. Fig.
\ref{fig:mt2mct2_uprising} shows the $\mct2$(uprisen) and
$\mt2$(laid) distributions of slepton pair production events, where
a pair of right(left) handed
sleptons($m_{\tilde{l}_{R(L)}}=191~(256)$ GeV) are produced(sps5),
and each slepton decays to $l + \tilde{\chi}_0^1$ with
$m_{\tilde{\chi}_0^1} = 119 $ GeV.
This is also a parton level result and we used $\chi =
m_{\tilde{\chi}_0^1}$. In the $\mt2$ distribution, there exists a BP
which is made by the endpoint of $\tilde{l}_{R}$ but buried in the
$\tilde{l}_{L}$ signature. One can see that such a BP is enhanced in
the $\mct2$ projection, where the $J_{max}(\chi)$ factor for
$\tilde{l}_{R}$ is 5.3 for $\chi = m_{\tilde{\chi}_0^1}$. \emph{With
this example, one might naturally think about the possibility to
resolve some mass hierarchies with reduced systematic uncertainties
using the $\mct2(\chi)$ projection of some inclusive event data.}
The mass hierarchies will be represented by several $\p0$ values,
for instance, here there are two $\p0$ values to be resolved as
follows
\begin{eqnarray}
|\p0^{(1,2)}|=\frac{m_{\tilde{l}_{(R,L)}}^2-m_{\lnp}^2}{2
m_{\tilde{l}_{(R,L)}}}, \label{eqslp0}
\end{eqnarray}
each of which can be extracted from the first and the 2nd endpoints
of the $\mct2(\chi)$ or $\mt2(\chi)$ distributions, respectively.

 As we will see in the next section, $\mct2(\chi)$ projection can be a
very powerful tool for resolving several $|\p0|$ values originated
from various mass hierarchies between new particles. Basically,
wherever there exist symmetric decay chains with various mass
hierarchies, one can try to resolve them in terms of various $|\p0|$
values. If the symmetric chain involves three body decays like
($2\times (Y\rightarrow qqX$)), the effective $|\p0|$ value which
contributes to the endpoint of $\mct2(\chi)$ or $\mt2(\chi)$ will be
$(m_Y-m_X)/2$. (Here the $\mct2(\chi)$ and $\mt2(\chi)$ is also
defined with only 2 visible particles with properly modified missing
$\etmiss$.) From now on, we define $|\p0|$ by \dis{
|\p0|&\equiv \textit{Maximal absolute momentum of daughter}\\
&\textit{particles in the rest frame of mother particle}\\
&=\left\{
\begin{array}{l}(m_Y^2-m_X^2)/(2 m_Y)\,\,\textit{for 2 body decay}\\
(m_Y-m_X)/2 \,\,\textit{for 3 body decay.}
\end{array}
\right.\label{eqp0def}
}
Therefore, the $\mct2(\chi)$ projection enables us to implement some
precision endpoint measurement using the inclusive signal and background
data with large uncertainties. The amplification of
some BPs by $\mct2(\chi)$ projection increases our observability for
meaningful signal endpoints in $\mt2$ distribution, which are buried
in severe backgrounds.

\section{Applications}\label{sec:mct2experiment}
In this section, we will describe more useful example of
$\mct2(\chi)$ projection in order to obtain some constraints between
$m_{\tilde{g}}$, $m_{\tilde{q}}$, and other superparticle masses in
a situation where large uncertainties are involved with jets.
 Specifically, an example of measuring some superparticle
mass constraints via $\mct2$ projection is shown, using 6 hard  jets
+ $\etmiss$ signal from a SUSY model. The mSUGRA type SUSY benchmark
spectrum is chosen in Table \ref{tab:susyspec}.
\begin{table}[h]
\begin{tabular}{|c | c | c | c |}
\hline
 $m_{\squark}$& $m_{\gluino}$ & $m_{\chargino/\2ndlnp}$ &
$m_{\lnp}$ \\
\hline $1026.3\sim1036.6\GeV$ & $649.4 \GeV$ & $182.1/181.2 \GeV$ &
$98.6 \GeV $\\
\hline
\end{tabular}
\caption{\label{tab:susyspec} A benchmark SUSY spectrum.
SOFTSUSY\cite{softsusy} is used to calculate the spectrum with
mSUGRA model parameters, $m_{1/2}=250
\GeV$,$M_0=900\GeV$,$\tan{\beta}=10$, $\mu>0$, and $A_0=0.$}
\end{table}

\begin{figure}[t]
 \begin{center}
\epsfig{figure=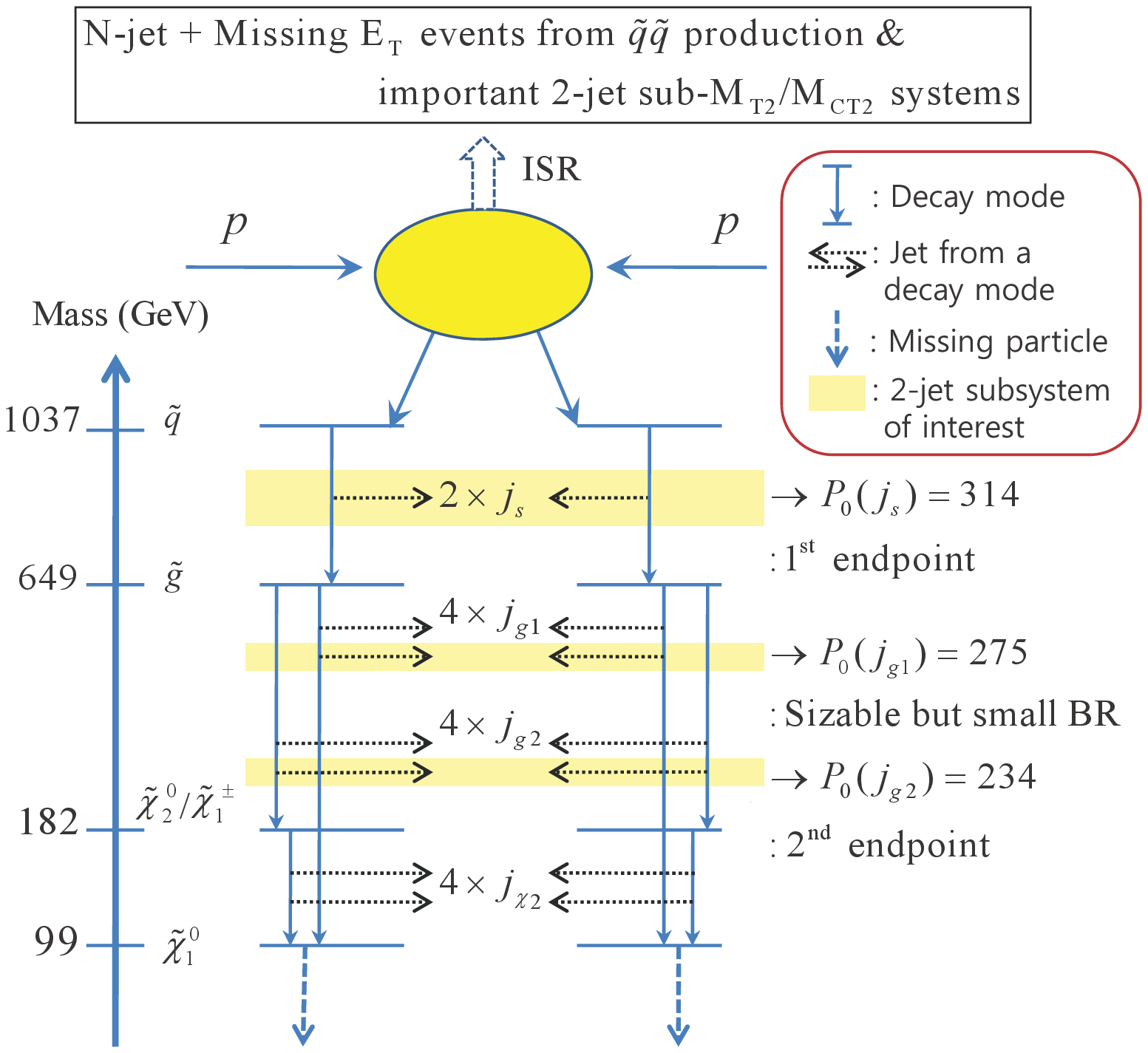,width=8.5cm,height=8cm}
\caption{A SUSY event topology for hard N($\geqq6$)-jets + $\etmiss$
signature and the important 2-jet Subsystem-$\mt2$ and $\mct2$
systems(yellow boxes). $j_s$ : a jet from $\squark$ decay to
$\gluino$,  $j_{g1}$ : a jet from $\gluino$ decays to $\lnp$,
$j_{g2}$ : a jet from $\gluino$ decay to $\2ndlnp/\chargino$,  and
$j_{\chi}$ : a jet from $\2ndlnp/\chargino$ decay to $\lnp$. }
\label{fig:susyevent}
\end{center}
\end{figure}

\noindent  At this benchmark point, squark and slepton masses are
heavier than gluino by a few hundreds $\GeV$. Also the 2nd
neutralino $\2ndlnp$ and chargino $\chargino$ are mostly wino so
that their masses are nearly degenerate. Thus, if we take into
account the production and decay rates at the LHC energy of 14 TeV, the important new particle mass
hierarchies can be categorized by 4 mass scales listed in Table
\ref{tab:susyspec}. $\tilde{g}\tilde{g}$ and $\tilde{g}\tilde{q}$
productions are the dominant superparticle generation processes with
$\sigma \sim 3 \pb$ for both cases. $\tilde{q}\tilde{q}$ production
also has a sizable cross section about $0.5 \pb$. Lepton production
is suppressed here because sleptons are also very heavy so that
leptons cannot be generated via some cascade decay chain of gluino
or squarks. So most of new physics signals are hard N-jets +
$\etmiss$ signal from squark or gluino decays. Several important
squark and gluino decay chains give the following branching ratios
\begin{enumerate}
    \item $BR(\squark\rightarrow \gluino+q)\sim 65\%$
    \item $BR(\gluino\rightarrow \chargino+q+ \bar q)\sim 44\%$
    \item $BR(\gluino\rightarrow \2ndlnp +q+ \bar q)\sim 25\%$
    \item $BR(\gluino\rightarrow \lnp +q+ \bar q)\sim 15\%$
\end{enumerate}
Here, all the listed gluino branching ratios are for 3 body decay
processes via virtual squarks. The inclusive hard N($\geq6$) jets +
$\etmiss$ signal can come from heavy squark pair production where
each squark decays to $j_s + \gluino$, and subsequently, $\gluino
\rightarrow j_g j_g+\lnp/\chargino/\2ndlnp$ through 3 body decay
chain,
\dis{
&pp \rightarrow G(-\delta_T) +\tilde{q}+\tilde{q}\\
&\quad\rightarrow G(-\delta_T) + 2\times (j_s+\tilde{g}(\rightarrow j_g
j_g\,\lnp/\chargino/\2ndlnp)),\\
&\chargino/\2ndlnp \rightarrow W/Z^{*}(\rightarrow
j_{\chi}j_{\chi})+\lnp,\label{eqsqproduction} } where $G(-\delta_T)$
is the extra particles(e.g. ISR) which gives a recoil transverse
momentum $\delta_T$ to the $\squark\squark$ system. Fig.
\ref{fig:susyevent} shows the event topology of the $\squark\squark$
production we consider. $j_s$ and $j_g$ in (\ref{eqsqproduction})
denote the jets from squark decay and gluino decay, respectively.
$j_g$ can be separated to $j_{g1}$ and $j_{g2}$ as in Fig.
\ref{fig:susyevent}, according to their decay processes. In this
paper, $\chargino$ and $\2ndlnp$ are assumed to decay to light
jets($j_{\chi}$) + $\lnp$ via $W$ or $Z$ bosons if they are produced
in the gluino decay.

In this regard, phenomenologically the most important question would
be, ``How can we measure the superparticle masses using N-jets +
$\etmiss$ signal in a situation with large jet uncertainties in
identifying true signal jets with complex combinatorics?" One might
attempt to measure the gluino or squark masses using the so-called
`$\mt2$-kink method \cite{mt2kink}'. According to this method, one
can have a good chance to measure the squark(or gluino) and $\lnp$
masses simultaniously by extracting the position of the kink, if  6
jets(or 4 jets) from squark(or gluino) pair decays are efficiently
identified. In particular, for gluino mass measurement, if squarks
are very heavy so that they are decoupled at the LHC, then most of
the hardest 4 jets would be from the decays from the gluino pair,
the efficiency of choosing the correct 4 jets can be quite nice.
However, when squarks are heavier than gluino by just a few hundred
$\GeV$, then the efficiency gets worse, particulary for
$P_T(j_s)\geq P_T(j_g)$ when ($m_{\squark}-m_{\gluino}$) is
comparable to, or larger than ($m_{\gluino}-m_{\lnp}$). Although the
squark pair production rate decreases as the
($m_{\squark}-m_{\gluino}$) increases, the two $j_s$ with hard $P_T$
are more likely to be selected in the gluino $\mt2$ calculation, and
badly pollute the endpoint structure \cite{mt2kinkinclusive}.
Furthermore, the ISR can be an important source of jet backgrounds
with hard $P_T$ also \cite{mt2kinkisr,qcdisr}. Thus, there exists a
large amount of systematic uncertainty in correctly estimating the
jet backgrounds of gluino $\mt2$.

As we will describe soon, the $\mct2$ projection can help resolving
several mass differences hidden in various inclusive jet signatures
in general. The mass differences are represented by several $|\p0|$
values in the endpoints measurements of the $\mt2$ and $\mct2$
distributions, as we discussed in the previous sections. Such
$|\p0|$ values are listed in Fig. \ref{fig:susyevent}, whose values
have been calculated by (\ref{eqp0def}).

Here we concentrate on measuring the mass difference between squark
and gluino since the usual gluino mass measurements have a lot of
uncertainties as we explained in the previous paragraph. After the
description of this specific example, we will present a
generalization of the $\mct2$ amplification for resolving several
mass hierarchies which have been buried in various inclusive
signatures, so regarded as meaningless in the $\mt2$ distribution.

Our strategy for measuring the squark-gluino mass difference is
naturally focusing on the two $j_s$ rather than trying to select
four $j_g$ from gluino pair decays. Surely, there exists a large
amount of uncertainty in correctly choosing two $j_s$ and estimating
the backgrounds also. However, we attempted to calculate
$\mct2(\chi)$ and $\mt2(\chi)$ for the Subsystem ($\squark\squark
\rightarrow j_s j_s \gluino\gluino$), considering two gluinos as the
effective missing particles for the system. Once we could construct
the Subsystem-$\mt2(\chi)$ (M. Burns et. al. in \cite{mt2ext}) using
the correct pair of two $j_s$, then in the limit of vanishing
$\delta_T$, its endpoint corresponds to $M^2$ in (\ref{eqmc4}) with
$|\p0|$ given by
\begin{eqnarray}
|\p0(j_s)|=\frac{m_{\squark}^2-m_{\gluino}^2}{2 m_{\squark}}=314.2
\GeV.\label{eqsubp0}
\end{eqnarray}
\begin{table*}[t]
\begin{tabular}{|c | c | c | c | c | c | c | c|}
\hline $\chi/|\p0(j_s)|$& $J_{max}(\chi)$ &
$\mt2^{max}(\chi)^{true}$ & $\mt2^{max}(\chi)^{exp}$ &
 $\mct2^{max}(\chi)^{true}$ & $\mct2^{max}(\chi)^{exp}$
 & $ \delta^{sys}_{\mct2^{max}}/\delta^{sys}_{\mt2^{max}}$
 & $ \delta^{sys}_{\p0}(\mct2)/\delta^{sys}_{\p0}(\mt2)$\\
\hline $1.24$ & 12.2 & $814.8 \GeV$ & $810.2 \pm 19.5 \pm
\textit{31.6}$
& $518.6 \GeV$& $520.5 \pm 1.6 \pm \textit{0.3}$ & \textit{0.01}& \textit{0.12}\\
$2.57$ & 3.9 &$1179.8 \GeV$ & $1150.2 \pm 18.3 \pm \textit{30.4}$
& $998.5 \GeV$ & $1003.1 \pm 4.5 \pm \textit{0.8}$& \textit{0.03} & \textit{0.11}\\
\hline
\end{tabular}
\caption{\label{tab:mct2mt2exp} Expected/Fitted endpoints of the
$\mct2$ and $\mt2$ for $\chi=1.24|\p0(j_s)|,2.57|\p0(j_s)|$.
For each measured endpoint value, the first error is
statistical and the second one is systematic in the fitting. One can find
that statistical errors are reduced by a factor of $1/J_{max}(\chi)$
as expected, while the the systematic errors are reduced by a factor
of $1/J_{max}^2(\chi)$, or less.}
\end{table*}
Therefore, the measurement of the Subsystem-$\mt2(\chi)$ endpoint
can provide us a constraint between $m_{\squark}$ and $m_{\gluino}$.
Once we assume that the $m_{\squark}$ and $m_{\lnp}$ masses could be
already obtained through the $\mt2$-kink methods using 6 hardest
$P_T$ jets, then $m_{\gluino}$ might be also measured by
(\ref{eqsubp0}). However, as commented previously, the endpoint
measurement of the Subsystem-$\mt2(\chi)$ distribution is not easy.
Choosing the correct pair of two $j_s$ suffers from the jet
uncertainties, and hence some effective event selection cuts or methods
purifying the signals should be applied. Thus, we describe another
approach to the endpoint measurement of the Subsystem-$\mt2(\chi)$.
In short, it can be summarized as follows :
\\

\textit{We do not care whatever backgrounds there exist. Once
there is a dim BP from a signal endpoint in the $\mt2$
distribution with large systematic uncertainties, then we can try to
extract the position of the amplified BP in the $\mct2$ projection.}
\\

With this point of view, we calculate the Subsystem-$\mct2(\chi)$
and Subsystem-$\mt2(\chi)$ variables for all possible 15 pairs of
the two jets among the 6 highest $P_T$ jets in an event with
N({$=6,7$)-jets + $\etmiss$. If the event is from a real
squark pair production, then there exists at least one true pair of
two $j_s$ among the 15 pairs of possible two jets, and it might
consistently contribute to the meaningful endpoint in the
Subsystem-$\mt2(\chi)$ distribution, although the slope change must
be faint. The portion of true two $j_s$ pair will be much more
reduced if we take into account the background events, e.g.
$\gluino\gluino$, $\gluino\squark$ + etc with hard ISR effects.
Anyway, using the events with N($=6,7$)-jets +
$\etmiss$, we tested the possibility to measure the corresponding
amplified BP in the $\mct2$ projection with less systematic errors.
Extracting meaningful endpoints using more inclusive events with
general N($\geq6$)-jets + $\etmiss$ will be discussed in the latter
part of this section and a forthcoming paper \cite{nextmct2}. We
found that even with this small fraction of N($=6,7$)-jets +
$\etmiss$ events, a sizable number of events survive the usual new
physics cuts, and contribute to some meaningful amplified BPs of the
inclusive Subsystem(IS)-$\mct2(\chi)$. To reconstruct the BPs in the
IS-$\mct2(\chi)$ and $\mt2(\chi)$, the 15 values of
Subsystem-$\mct2(\chi)$ and Subsystem-$\mt2(\chi)$ are constructed
with given trial gluino masses $\chi$ for an event as follows and we
made histograms of all the calculated values without any
preferential jet selection among the 6 hardest jets,
\begin{eqnarray}
\mctt2^{(n)}(\chi) &=& \min_{\sum{{\bf k}_{iT}}=\bfetmiss^{(n)}}
[\max{\{\mmctt^{(n,1)},\mmctt^{(n,2)}\}}]\label{eqmct22}\\
\mmctt^{(n,i)}(\chi)^{2} &=& \chi^2 + m_{i}^{(n) 2} + 2 (e_{i}^{(n)}
e_{\chi_{i}} \pm {\bf p}_{iT}^{(n)} \cdot {\bf
k}_{iT}) \nonumber\\
\bfetmiss^{(n)} & \equiv & \bfetmiss + {\bf q}_{T}^{(n)}. \nonumber
\end{eqnarray}
The index $n(=1, \cdots, 15)$ means a specific combination to select
two $j_s$ among the 6 hardest $P_T$ jets. ${\bf p}_{iT}^{(n)}$ are
the transverse momenta of the two jets selected as $j_s$ in the lab.
frame. $\bfetmiss$ is the total missing transverse momenta, and
$\bfetmiss^{(n)}$ is modified missing transverse momenta for the
subsystem with ${\bf q}_{T}^{(n)}$ indicating the ${\bf p}_T$ sum of
the remaining 4 jets which are selected as four $j_g$ in the $n$-th
combination. $\chi$ is meant to be a trial gluino mass with this
setup of visible and missing momenta.
\\

We simulated the $\squark\squark,\gluino\gluino,\squark\gluino$
production events of $10 \invfb$ at the LHC energy of 14 TeV
using PYTHIA 6.4 \cite{pythia} with ISR/FSR turned on. Fully
showered and hadronized events were passed to the PGS 4.0 detector
simulator \cite{pgs}. The energy resolution parameter in the
hadronic calorimeter was given by $\Delta E/E = 0.6/\sqrt{E}$, and
jets were defined using a cone algorithm with $\Delta R = 0.5$.
To be more reliable for the new physics event
experiments, we imposed several cuts for pure N-jets + $\etmiss$
events \cite{atlas,cms,alpha} although we did not include the SM
background event samples for simplicity.

The event selection cuts were as follows:
\begin{enumerate}
    \item No leptons, no $b$ jets in the event,
    \item Number of jets $=6,7$ with $P_{T}^{1st,2nd}\geq 100\GeV,\\
          P_{T}^{6th} \geq 50 \GeV$,
    \item $\etmiss \geq 100\GeV$,
    \item $\alpha^{n} \geq 0.45$ with $n(=1,\cdots, 15)$ for the pairs\\
     of selected two jets,
    \item $\Delta_T(\equiv|{\bfetmiss}+{\sum}_{j=1,\cdots, 6}{\bf p}_T^j|)\leq 30 \GeV$,
\end{enumerate}

\noindent where the index-$j$ denotes the 6 hardest jets.
The $\alpha^{n}\equiv P_{T2}^{n}/m_{jj}^{n}$
\cite{alpha} where $P_{T2}^{n}$ and $m_{jj}^{n}$ are the $P_T$ of
the second hardest jet and dijet invariant mass, respectively, for the
$n(=1,\cdots,15)$-th pair of selected two jets. If $\gluino\gluino$ decay
directly to $\lnp\lnp$, then  $\Delta_T$ becomes $\delta_T$ which
is the total transverse momentum of the squark pair system defined
in (\ref{eqsqproduction}). As explained in the previous section, the
fifth cut is imposed to suppress the shift of the $\mct2(\chi)$
endpoint (\ref{eqmct2max}) under arbitrary transverse boost of two
mother particle system($\squark\squark$).\footnote{The $\mct2$
endpoint with a sizable $\delta_T$ will be studied in a forthcoming
paper \cite{nextmct2}.} We calculated the $\mct2^{(n)}(\chi)$ and
$\mt2^{(n)}(\chi)$ for all the events passing the cuts. The single
jet invariant masses, $m_{i}^{(n)}$ in calculating the $\mt2^{(n)}$
and $\mct2^{(n)}$ of (\ref{eqmct22}) was ignored since we found that
it is quite helpful for the endpoints of both variables to be
located at the expected position, reducing the jet energy resolution
effect. Ignoring the jet masses also satisfies all the inequalities
of $\mt2$ and $\mct2$ and the constructed variables can saturate to
the boundary value because of enough statistics of light QCD jets in
the events. In our SUSY event sample, the jet multiplicity ratio is $\frac{\sigma(N_{jet}=6,7)}{\sigma(N_{jet}\geq6)}\sim 0.23$. With the relatively small portion of the event
sample ($N_{jet}=6,7$) surviving the cuts, we still could observe
the BP in the IS-$\mt2(\chi)$, and the amplified BP in the
IS-$\mct2(\chi)$ also.

Fig. \ref{fig:fig4}(a,b)(two histograms in the top) shows the $\mt2$
and the corresponding $\mct2$ distributions for $\chi = 1.24
|\p0(j_s)| = 389.7 $ GeV. Fig. \ref{fig:fig4}(c,d)(two histograms in
the bottom) are also $\mt2$ and $\mct2$ for $\chi = 2.57 |\p0| =
806.5 $ GeV. The endpoint enhancement factor, $J_{max}$ is 12.2 for
(a)$\rightarrow$(b), and 3.9 for (c)$\rightarrow$(d). We constructed
many histograms with respect to various $\chi$ values. Among them,
we selected the two cases as shown in Fig. \ref{fig:fig4}. Actually,
compared to $|\p0|$, a much larger or smaller value of $\chi$ is not
a good choice as explained in the previous section. Based on our
trial experience, choosing $\chi$ of order of $|\p0|$ shows a proper
amplification feature with clean breakpoint structures, and this
selection can be realized by observing an average value of $|P_T|$
of the hardest jets. A strong $\Delta_T$-suppression cut was quite
effective for observing a BP with a small slope discontinuity at the
expected endpoints in the $\mt2$ distributions, and we could observe
the amplified BPs also at the expected position in the corresponding
$\mct2$ distributions. Actually, this cut is effective for selecting
the $\squark\squark$ production event. The expected endpoints are
pinpointed by a red dashed line in each figure. The bin interval was
selected as the best one among several bin candidates,
by which the histogram near the expected BP regions shows its
characteristic feature maximally while keeping its statistical
relevance.

In Fig. \ref{fig:fig4}(a), the $\mt2(\chi=389.7)$ distribution shows
a small slope discontinuity near the expected endpoint $814.8 \GeV$.
One can find that the faint BP structure is amplified near the
expected endpoint $518.6\GeV$ in the $\mct2(\chi=389.7)$ (Fig.
\ref{fig:fig4}(b)) projection, producing a sharp cliff wall.
Similarly, there exists a small BP near the expected endpoint
$1179.8$ GeV of $\mt2(\chi=806.5)$ distribution in
Fig.\ref{fig:fig4}(c), and the BP is transformed to another BP with
a sharp cliff near the expected endpoint $998.5\GeV$ of
$\mct2(\chi=806.5)$ in Fig. \ref{fig:fig4}(d).

Table \ref{tab:mct2mt2exp} shows the expected and fitted endpoints
of the $\mt2$ and $\mct2$ distributions for $\chi=1.24|\p0|$ and
$2.57|\p0|$. The errors after the fitting process are also listed in
the columns of $\mt2^{max}(\chi)^{exp}$ and
$\mct2^{max}(\chi)^{exp}$. Here, the first error represents statistical
one, and the second represents the systematic one in the fitting
process. The fit model functions for the endpoints were a Gaussian
smeared linear function for a signal, and a linear function for the
backgrounds in the $\mt2$ distributions (Fig. \ref{fig:fig4}(a,c)),
\begin{eqnarray}
f(m)&=&\Theta(m-\mt2^{max})\frac{1}{\sqrt{2\pi}\sigma}\int^{\mt2^{max}}
e^{(-\frac{(m-m')^2}{2\sigma^2})}\nonumber\\
&\times& a_1(m'-\mt2^{max})dm' + a_2 m + a_3.\label{fit1}
\end{eqnarray}
Also, the endpoint region of the $\mct2$ distribution
(Fig. \ref{fig:fig4}(b,d)) was fitted by the Gaussian smeared step
functions for signals, and step function for backgrounds,
\begin{eqnarray}
f(m)&=&\frac{a_1\Theta(m-\mct2^{max(1)})}{\sqrt{2\pi}\sigma}
\int^{\mct2^{max(1)}}
e^{(-\frac{(m-m')^2}{2\sigma^2})}dm'\nonumber\\
&+&\frac{a_2\Theta(m-\mct2^{max(2)})}{\sqrt{2\pi}\sigma}\int^{\mct2^{max(2)}}
e^{(-\frac{(m-m'')^2}{2\sigma^2})}dm''\nonumber\\
&+& a_3.\label{fit2}
\end{eqnarray}
Fitting was implemented by the binned $\chi^2$ method using MINUIT
with MINOS error processing \cite{minuit} which takes into account
both parameter correlations and non-linearities. The two step
functions for signal endpoints was designed in (\ref{fit2}) to fit
two cliffs appearing in the $\mct2$ distributions. We will explain
the second endpoint below the 1st endpoint which were pointed by red
dashed line in the $\mct2$ distribution (Fig. \ref{fig:fig4}(b,d))
in the latter part of this Section. For simplicity, we used the same
smearing widths $\sigma$ for both the two endpoint cliffs in the
$\mct2$ distribution. Actually, the uncertainty from the endpoint
broadening effect by some widths is not our issue here. As described
in Sec. \ref{sec:mct}, it is simply because the endpoint uncertainty
reduction factor is expected to be just the order of $1/J_{max}$ in
that case, and the reduced width is compensated by the explicit
error propagation factor $J_{max}$ for obtaining $|\p0|$ value.
It is natural that there is no advantage in using
$\mct2$ projection for reducing the statistical errors. Anyway, the
fitted endpoint values were found not to be very sensitive to the
choice of $\sigma$ in the range $0-10 $ GeV for $\mt2$, and 0--1 GeV
for $\mct2$, and we fixed the smearing widths $\sigma$ to be 5 GeV
for the $\mt2$ endpoint, and 0.5 GeV for the $\mct2$ endpoints. The
width for the $\mct2$ endpoint is a suppressed value from the width
of the $\mt2$ endpoint by $O(1/J_{max})$.
\begin{figure*}[ht]
 \begin{center}
\epsfig{figure=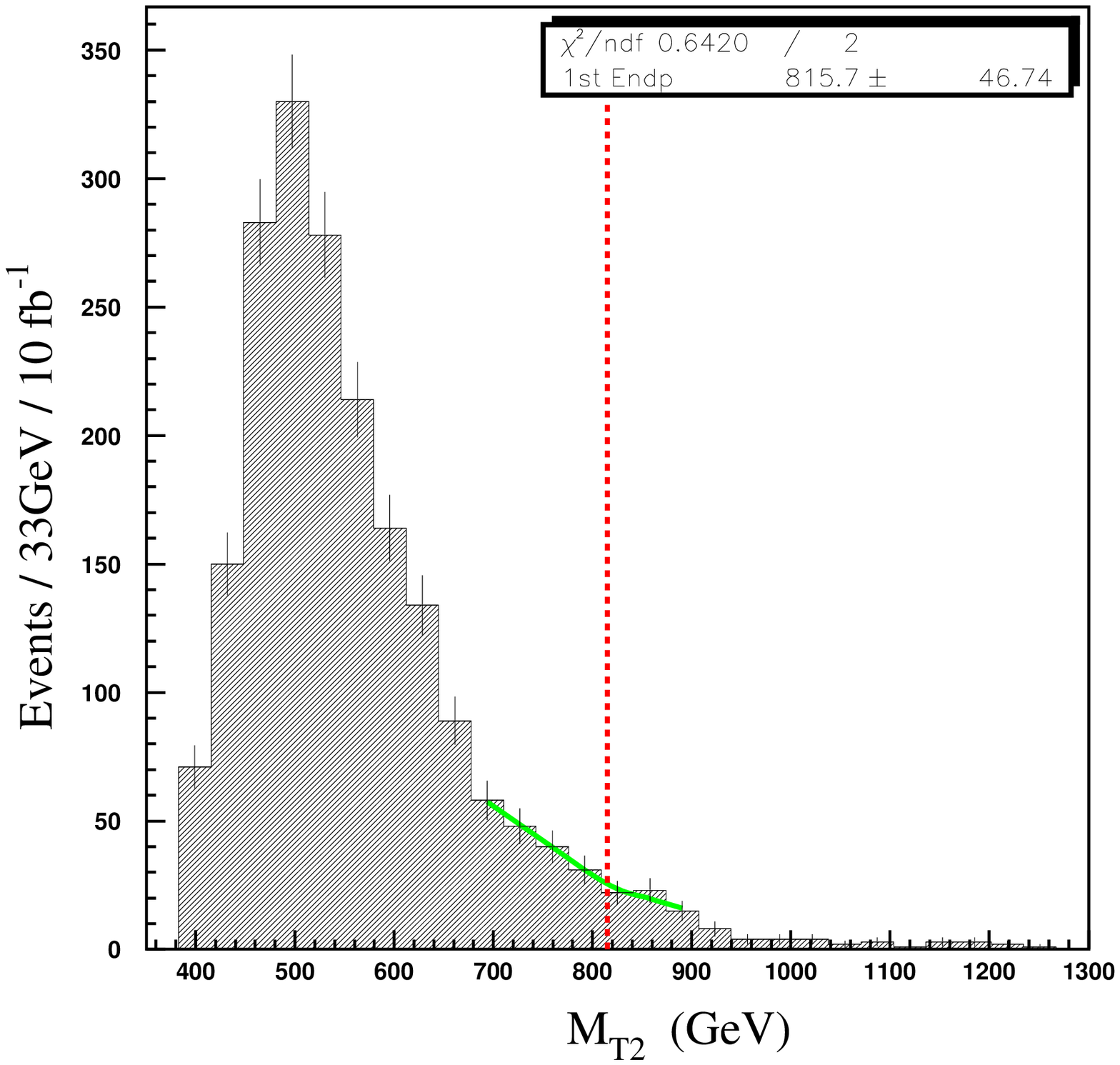,scale=0.43}
\epsfig{figure=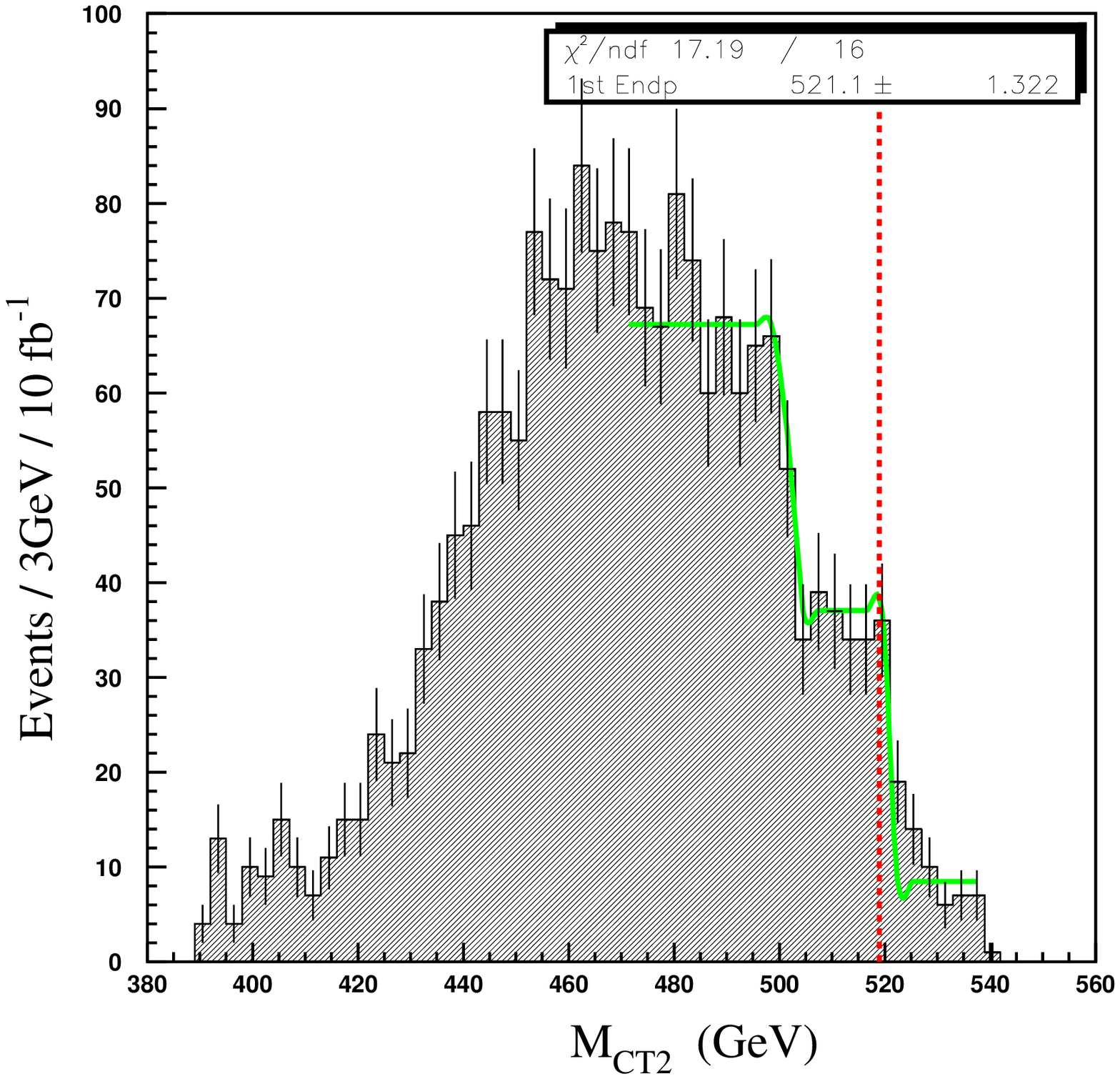,scale=0.43}
\epsfig{figure=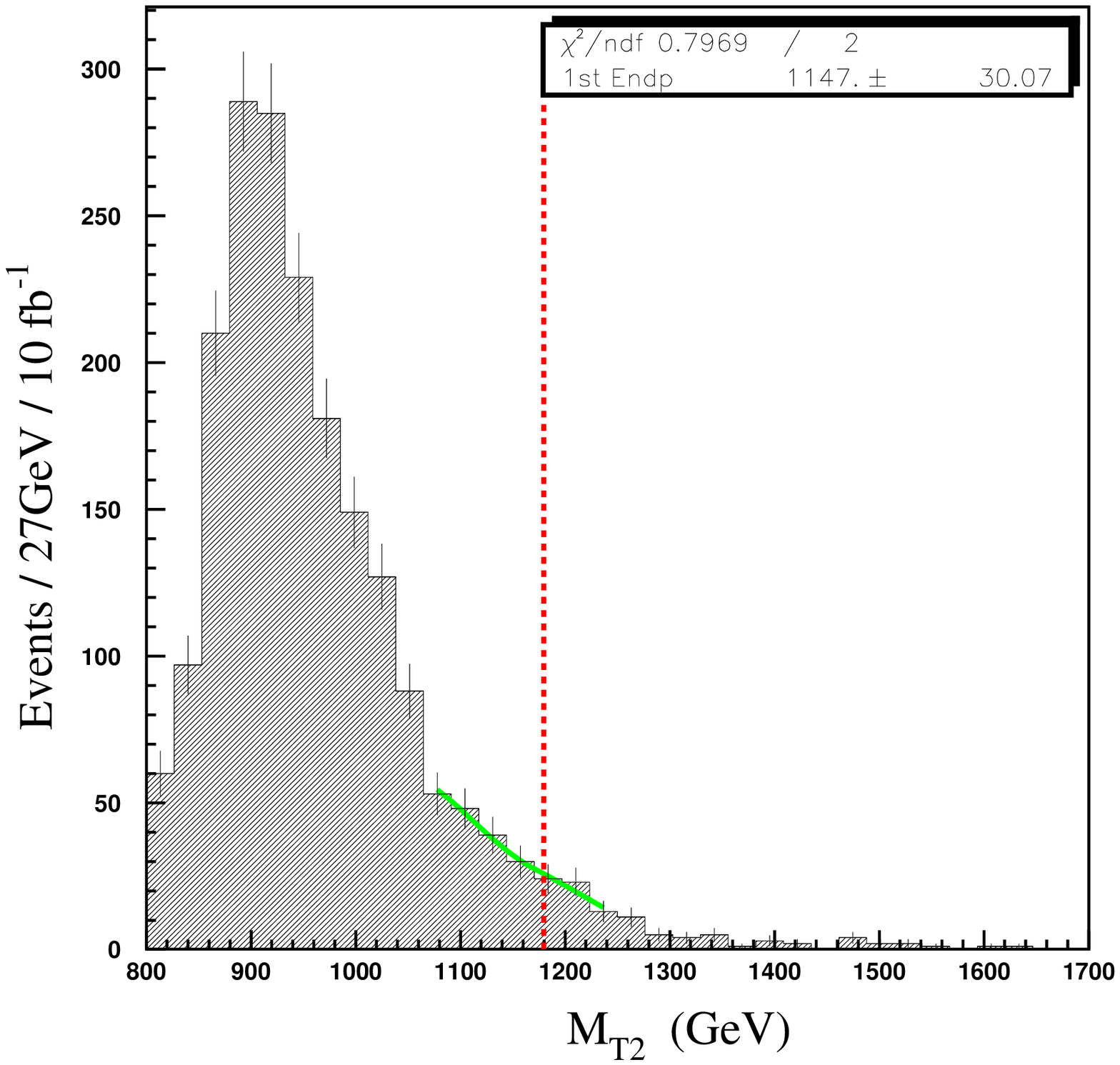,scale=0.43}
\epsfig{figure=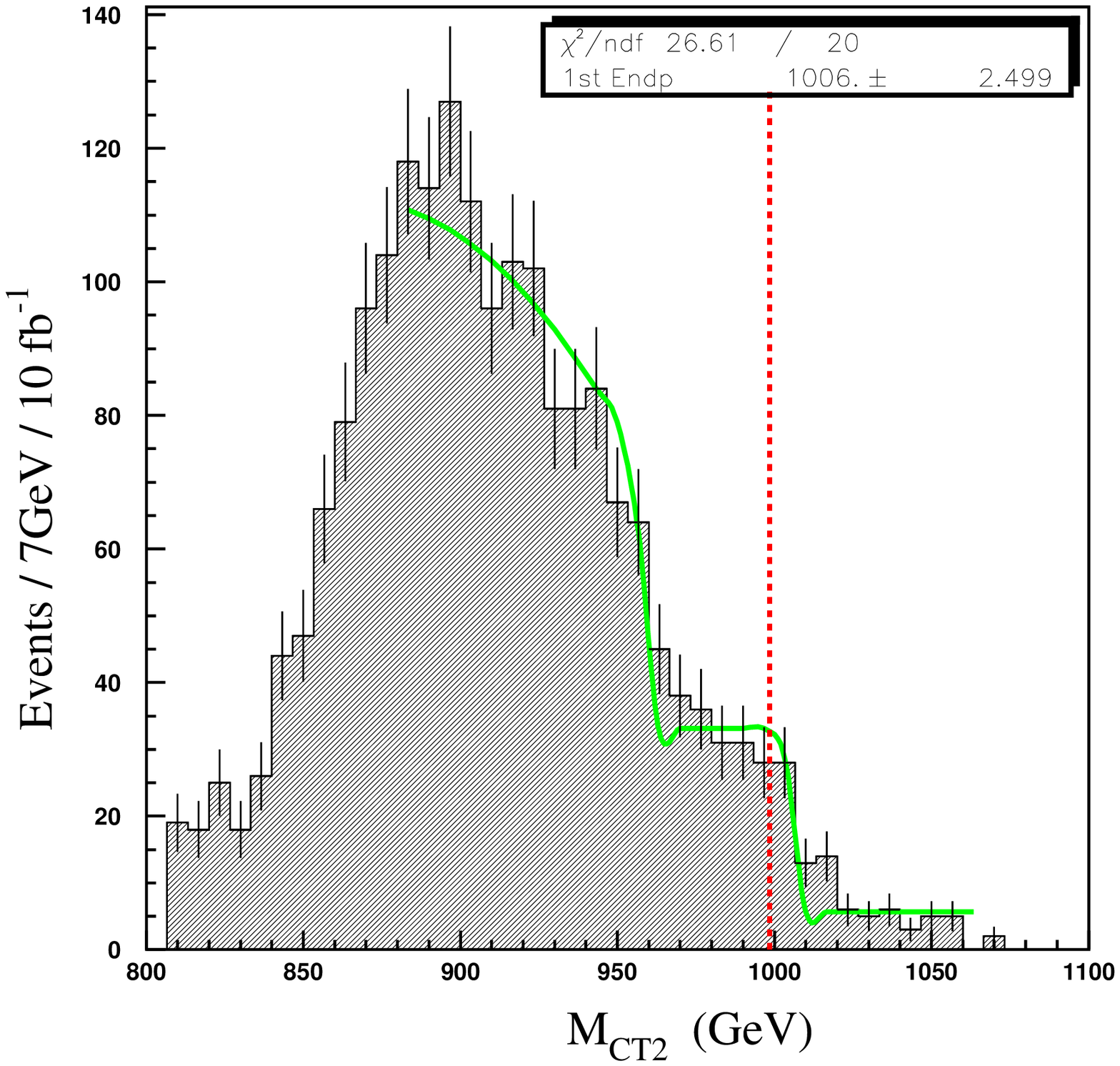,scale=0.43}
 \caption{Endpoint fittings
of the $\mt2$ (left columns (a) and (c)) and $\mct2$ (right columns
(b) and (d)) distributions
for  $\chi=1.24 |\p0(j_s)|$ ((a) and (b), top row), and for $\chi =2.57 |\p0(j_s)|$ ((c) and (d), bottom row).  The green lines are  the fitted model functions and the red dashed lines indicate the expected endpoints :\\
$\mt2^{max}(1.24|\p0|)=814.8 \GeV$, $\mct2^{max}(1.24|\p0|)=518.6
\GeV$, $\mt2^{max}(2.57|\p0|)=1179.8 \GeV$,
$\mct2^{max}(2.57|\p0|)=998.5 \GeV$. } \label{fig:fig4}
\end{center}
\end{figure*}

The green lines in Fig. \ref{fig:fig4} represent the fitted model
function. In order to estimate the systematic uncertainty, we
attempted to fit various endpoint regions with the simple model
functions while keeping the $\chi^2/ndf < 2$ after fitting( $ndf$ =
the number of fitted bins -- 1). The small value of $\chi^2/ndf$
might guarantees the plausibility of the fit range described by the
model function. Actually, for anyone trying to find some optimal
range of the fitting in the $\mt2$ distribution,
 he easily notices that there exist a large number of
choices for a lower boundary and an upper boundary, which is
consistent with  small $\chi^2/ndf$ after fitting. In this
situation, because the BP is quite a faint one can observe that the
fitted endpoint parameter changes very much with a large uncertainty as
the fit range of validity is varied. On the other hand, the fitting
of the $\mct2$ endpoint regions can be different. Due to the
emergence of the sharp cliffs, which are the amplified BPs, we could
clearly elaborate our effective fit model functions and the fit
ranges of validity. Actually, there is another meaningful BP near
$690$ GeV in the $\mt2(\chi=1.24|\p0|)$ distribution, which is now
clearly seen as the cliff near $505$ GeV in the $\mct2$
projection(For $\chi=2.57|p0|$, a similar BP enhancement is
observed). It looks even more clearer than the first BP we
expected for the squark pair decays. In the past, one would have neglected
such a BP buried in the region of the low $\mt2$ distribution, as
well as suffer from large systematic uncertainties in extracting the
position of the first BP. Now, in addition to the first endpoint(red
dashed), we can try to extract the amplified second lower endpoint and
interpret its origin with less uncertainty.

One can easily see that the second endpoint is originating from the
mass differences between the $\gluino$ and $\2ndlnp,\chargino$,
represented by another $|\p0|$ value,
\begin{eqnarray}
|\p0(j_{g2})|=\frac{m_{\gluino}-m_{\2ndlnp,\chargino}}{2}
\simeq234\GeV.\label{eqp02}
\end{eqnarray}
\begin{figure*}[!ht]
 \begin{center}
\epsfig{figure=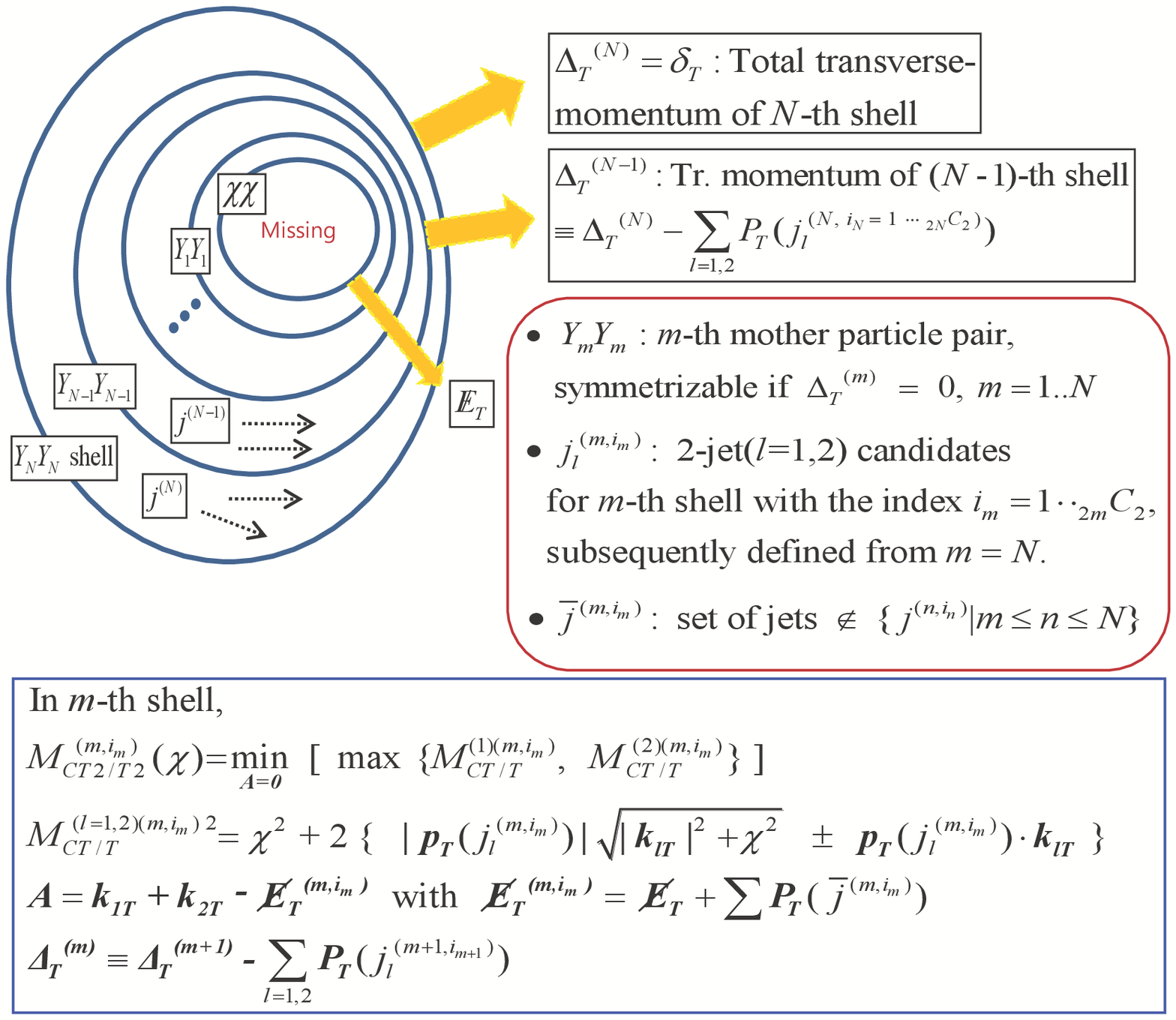,height=11.9cm,width=13.5cm}
\caption{Extension of the Subsystem-$\mct2$,$\mt2$ in
(\ref{eqmct22}) for $m$-th($m=1..N$) symmetrizable mother pairs with
symmetrization cut parameter $\Delta_T^{m}$ using inclusive hard
$n$($\geq2N$)-jet + $\etmiss$ signatures.}\label{fig:fig6}
\end{center}
\end{figure*}
This is the maximal momentum of jets from a gluino three body decays
to $j_gj_g + \2ndlnp/\chargino$ in the rest frame of the gluino.
Since the BR($\gluino \rightarrow j_{g2}j_{g2} + \2ndlnp/\chargino$)
is $4.6$ times larger than the BR($\gluino \rightarrow j_{g1}j_{g1}
+ \lnp$), gluino decays to $\2ndlnp/\chargino$ can contribute significantly to the inclusive N($\geq6$)-jets + $\etmiss$ signature
than the gluino decays to $\lnp$ since most of the
$\2ndlnp/\chargino$ hadronically decays to 2-jets($j_{\chi}$) +
$\lnp$ via $Z^*/W$ bosons. The point we want to emphasize here is
that, once there is some possibility of a pair of mother particles
to be symmetric in the lab. frame($\sum P_T= 0$), then as analyzed
in Sec. \ref{sec:mct}, it might contribute to some other endpoint in
the $\mt2(\chi)$ distribution with some $|\p0|$ value defined in
(\ref{eqp0def}). For example, in the squark pair production events
we consider, the symmetric condition of a pair of gluinos can be
realized if
\begin{eqnarray}
\delta_T - \sum P_T(j_s) = 0.\label{eqsymcon}
\end{eqnarray}
However, regardless of the gluino pair symmetry condition,
$\mt2^{(n)}$ and $\mct2^{(n)}$ in (\ref{eqmct22}) can include
Subsystem-$\mt2$ and $\mct2$ for the subsystem which consist of two
visible $j_{g2}^{(n)}$ + effective $\etmiss$, if the two ${\bf
p}_{iT}^{(n)}$ in (\ref{eqmct22}) correspond to the transverse
momenta of two $j_{g2}^{(n)}$. Here $j_{g2}^{(n)}$ means the 2
gluino jets $j_{g2}$, each of which comes from each mother gluino,
and is selected for ${\bf p}_{iT}^{(n)}$. The corresponding
subsystem is described in the third yellow box of Fig.
\ref{fig:susyevent}. If $j_{g2}^{(\bar{n})}$ denotes the other two
$j_{g2}$ which are not selected as the subsystem visible particles,
then the effective missing transverse momenta, $\bfetmiss^{(n)}$ in
the (\ref{eqmct22}), and $\Delta_T$ in the event selection cuts
become as follows:
\begin{eqnarray}
\bfetmiss^{(n)} & \equiv & \bfetmiss + {\bf q}_{T}^{(n)} \nonumber\\
&=& - \{\sum P_T(j_s) + \sum P_T(j_{g2}^{(n)}) + \sum
P_T(j_{g2}^{(\bar{n})})\nonumber\\
&& + \sum P_T(j_{\chi}) - \delta_T \}\nonumber\\
&&+\{\sum P_T(j_s)+\sum P_T(j_{g2}^{(\bar{n})})\}\nonumber \\
&=& - \sum P_T(j_{g2}^{(n)}) - \sum P_T(j_{\chi}) + \delta_T,\\
\Delta_T &\equiv& |\bfetmiss+{\sum}_{j=1..6}{\bf p}_T^j|\nonumber\\
&=& |\delta_T -\sum P_T(j_{\chi})|,\label{eqdeltat}
\end{eqnarray}
where the assumption that $j_{\chi}$ is not included in the 6
hardest $P_T$ jet candidates is adopted. Then, our strong $\Delta_T$
suppression cut implies that the effective missing transverse momenta are
\begin{eqnarray}
\bfetmiss^{(n)}\sim- \sum P_T(j_{g2}^{(n)}).
\end{eqnarray}
This indicates that Subsystem-$\mt2$ and $\mct2$ in (\ref{eqmct22})
can have a good chance of contributing to the well-defined endpoints
described by $|\p0(j_{g2})|$ value if the gluino symmetric condition
(\ref{eqsymcon}) is satisfied. In that case, the contribution
consistently forms the slight slope change in the $\mt2$
distribution, and it will emerge as a sharp cliff in the $\mct2$
distribution. In our example, the strong $\Delta_T$ suppression cuts
of (\ref{eqdeltat}) effectively acts as a $\delta_T$ suppression
cut, and then it will create a bias to choose small $\sum P_T(j_s)$
events because $\sum P_T(j_s)$ distribution is expected to be an
even convex function when $\delta_T$ is zero. Then, the symmetric
condition can be effectively satisfied for a lot of events, and
there exist large contributions to form some slope change near the
expected exact second endpoint.

However, one can easily notice that the $\Delta_T$ suppression cut
cannot guarantee the exact symmetric condition of the gluino pair.
In order to access to each symmetric mother particle pairs possibly
hidden in the N-decay shells, one can extend the definitions of
Subsystem-$\mt2$ and $\mct2$ of (\ref{eqmct22}) as shown in Fig.
\ref{fig:fig6}. Here, with hard $n$($2N\leq n \leq 4N$)-jet +
$\etmiss$ inclusive signatures, one can subsequently construct all
the possible $m$-th Subsystem-$\mct2^{(m)}$,$\mt2^{(m)}$ variables
from the highest shell, which has a well-defined generalized
symmetric condition parameter, $\Delta_T^{m}$. In general,
Subsystem-$\mct2^{(m)}$ and $\mt2^{(m)}$
 have $i_m^{max}$(=$_{2m}C_2$) multiplicity which is the number of jet combinations possible in the
$m$-th shell. As a result one can explicitly control those cut
parameters to observe the amplified BPs from several symmetrized
mother particle pairs. Furthermore, one can find that to realize the
hidden endpoints the one-dimensional decomposed version
\cite{decompose} of Subsystem-$\mct2^{(m)}$ and $\mt2^{(m)}$,
$M^{(m)}_{CT2\perp}$ and $M^{(m)}_{T2\perp}$, can be useful, if it
is constructed using only partial components of $ P_T(j^{(m)})$
perpendicular to $\Delta_T^{(m)}$ direction. In this regard, a more
general $\mct2$ spectroscopy will be discussed in a forthcoming
paper \cite{nextmct2}. As a result several mass hierarchies in
various new physics models can be probed by general
Subsystem-$\mct2(\chi)$ amplification using the inclusive
N($\geq2,4,6,\cdots$)-jets/leptons/photons + $\etmiss$ signatures,
with less systematic uncertainties.

Finally, we summarize the fit results of the first endpoints(red
dashed) in the column, $\mt2^{max}(\chi)^{exp}$ and
$\mct2^{max}(\chi)^{exp}$, of Table \ref{tab:mct2mt2exp}. Using
dozens of fit results for various fit ranges, we evaluated the
statistical errors as $19.5$ GeV for
$\mt2(\chi=1.24|\p0|)$, and 1.6 GeV for $\mct2(\chi=1.24|\p0|)$. The
listed systematic uncertainties were obtained by $1\sigma$ from the
fitted endpoint values. It is noticeable that the systematic errors
are suppressed in the $\mct2^{max}(\chi)^{exp}$ by the factor of
$1/J_{max}^2$ for $\chi=1.24|\p0|$ or less than that for
$\chi=2.57|\p0|$, while all of the statistical errors are reduced by
the order of $1/J_{max}$. As a result, the precision to measure the
$|\p0|$ can be enhanced at least by a factor of $J_{max}$ as listed
in the last column of Table \ref{tab:mct2mt2exp}. These results are
well matched with our expectation of the endpoint structure
enhancement by the $\mct2$ projections.

Including more various systematic error effects (the SM backgrounds,
the signal and BGs in higher orders, PDF) is beyond the scope of
this paper. But, the main point of this paper is clear: The features
of the BP enhancement by $\mct2$ projection has a more stronger
power in resolving the buried mass hierarchies as our situation gets
worse with large uncertainties.

\section{Conclusion }\label{sec:Conclusion}
We have introduced the {\it constransverse mass}, $\mct2$, and have
shown that it can significantly increase our ability to measure
the endpoints precisely by amplifying the small slope discontinuity of the endpoint region, which are usually buried in the complex backgrounds
with large uncertainties. As a result, the proposed $\mct2$ variable
for various subsystems with two visible particles can resolve the mass
hierarchies hidden in various inclusive signatures for the event
type, N($\geq2,4,6,\cdots$)-jets/leptons/photons + $\etmiss$. We hope
that this variable can be a useful tool in the LHC era.\\

\vskip 0.5cm
\acknowledgments{ WSC thanks Kiwoon Choi for useful discussions,
and Mihoko Nojiri for inviting him to the IPMU Workshop where this topic has been presented. WSC and JEK are supported in part by the Korea
Research Foundation, Grant No. KRF-2005-084-C00001, and  JHK is supported in part by Grant No. KRF-2008-313-C00162 and the FPRD of the BK21 program.}


\end{document}